\documentclass[12pt]{article}
\usepackage[font=small,labelfont=bf]{caption}
\usepackage{subcaption}
\usepackage{amscd,amssymb,amsxtra,amsthm,bm}
\usepackage{mathrsfs}
\DeclareSymbolFontAlphabet{\mathrsfs}{rsfs}
\usepackage[mathscr]{eucal}
\usepackage{mathabx}
\usepackage{savesym}
\usepackage[nosort]{cite}
\usepackage{wasysym}
\usepackage{amscd}
\usepackage{graphicx}
\usepackage{pifont}
\usepackage{float}
\usepackage{multicol}
\usepackage{amsfonts}
\usepackage{picture}
\usepackage{amssymb}
\usepackage{mathtools}
\savesymbol{iint}
\savesymbol{iiint}
\usepackage{amsmath}
\restoresymbol{TXF}{iint}
\restoresymbol{TXF}{iiint}
\usepackage{color}
\usepackage{clay}
\usepackage{hyperref}
\usepackage{slashed}
\hypersetup{colorlinks=true}
\hypersetup{linkcolor=black}
\hypersetup{citecolor=black}
\hypersetup{urlcolor=black}
\usepackage{setspace}
\usepackage{multirow}
\usepackage{verbatim}
\usepackage{accents}
\usepackage{placeins}
\usepackage{enumitem}
\usepackage[all]{xy}
\definecolor{refkey}{rgb}{0,.8,.2}\definecolor{labelkey}{rgb}{1,0,0}
\numberwithin{equation}{section}

\newcommand{\nn}{\nonumber}

\newenvironment{nalign}{
	\begin{equation}
	\begin{aligned}
}{
	\end{aligned}
	\end{equation}
	\ignorespacesafterend
}

\DeclareMathOperator{\Ima}{im}
\def\Spin{\text{Spin}}

\def\Tr{\text{Tr}\,}

\def\Z{\mathbb{Z}}

\def\C{\mathbb{C}}

\usepackage{array}
\newcolumntype{C}{>{$}c<{$}}

\begin{document}

\institution{DAMTP}{\centerline{${}^{1}$DAMTP, University of Cambridge, Wilberforce Road, Cambridge, 
CB3 0WA, United Kingdom}}
\institution{Cav}{\centerline{${}^{2}$Cavendish Laboratory, University of Cambridge, J.~J.~Thomson Ave, Cambridge, UK}}

\title{Global anomalies in the Standard Model(s) and Beyond}

\authors{Joe Davighi,\worksat{\DAMTP}\footnote{E-mail: {\tt jed60@cam.ac.uk}}  Ben Gripaios,\worksat{\DAMTP,}\worksat{\Cav}\footnote{E-mail: {\tt gripaios@hep.phy.cam.ac.uk}} and Nakarin Lohitsiri\worksat{\DAMTP}\footnote{E-mail: {\tt nl313@cam.ac.uk }}}

\abstract{
We analyse global anomalies and related constraints in the Standard Model (SM) and various Beyond the Standard Model (BSM) theories. We begin by considering four distinct, but equally valid, versions of the SM, in which the gauge group is taken to be $G=G_{\text{SM}}/\Gamma_n$, with $G_{\text{SM}}=SU(3)\times SU(2) \times U(1)$ and $\Gamma_n$ isomorphic to $\Z/n$ where $n\in\left\{1,2,3,6\right\}$. In addition to deriving constraints on the hypercharges of fields transforming in arbitrary representations of the $SU(3)\times SU(2)$ factor, we study the possibility of global anomalies in theories with these gauge groups by computing the bordism groups $\Omega^{\text{Spin}}_5(BG)$ using the Atiyah-Hirzebruch spectral sequence. In two cases we show that there are no global anomalies beyond the Witten anomaly, while in the other cases we show that there are no global anomalies at all, illustrating the subtle interplay between local and global anomalies. While freedom from global anomalies has been previously shown for the specific fermion content of the SM by embedding the SM in an anomaly-free $SU(5)$ GUT, our results here remain true when the SM fermion content is extended arbitrarily.

Going beyond the SM gauge groups, we show that there are no new global anomalies in extensions of the (usual) SM gauge group by $U(1)^m$ for any integer $m$, which correspond to phenomenologically well-motivated BSM theories featuring multiple $Z^\prime$ bosons.
Nor do we find any new global anomalies in various grand unified theories, including Pati-Salam and trinification models. %We do, however, show that there is a potential global anomaly in a five-dimensional grand unified theory based on $SO(18)$. 
We also consider global anomalies in a family of theories with gauge group $SU(N)\times Sp(M)\times U(1)$, which share the phase structure of the SM for certain $(N,M)$. Lastly, we discuss a BSM theory in which the SM fermions are defined using a spin$_c$ structure, for example by gauging $B-L$. Such a theory may be extended to all orientable four-manifolds, and we find no global anomalies.
}

\maketitle

\setcounter{tocdepth}{3}
\tableofcontents

\section{Introduction}\label{sec:intro}
The Standard Model (SM) has been tremendously successful in explaining all the data collected from collider physics experiments such as at the LHC, with the gauge, flavour, and Higgs sectors having been tested at the per mille, per cent, and ten per cent levels respectively~\cite{PhysRevD.98.030001}.  However, despite its successes, there are a number of unsolved problems in the SM. Some of these are experimental or observational in origin, such as the inability to account for the dark matter and dark energy that are observed by astrophysicists and cosmologists, while other problems appear to be more theoretical or aesthetic, such as the inability to describe physics beyond the Planck scale, and the (two) hierarchy problems associated with the two super-renormalisable operators in the SM lagrangian. It is clear that in order to offer a complete description of Nature, one must go beyond the Standard Model (BSM). 
In order to be a consistent quantum field theory, any BSM theory that we construct (as well as the SM itself)  must not suffer from any anomalies associated with its gauge group.

In fact, before we consider going beyond the SM, it is important to emphasise that there is not even {\em an} unique SM, but many possible Standard Model{\em s}, all of which are consistent with the same experimental data. The experimentally-observed SM gauge bosons and their interactions, together with the representations of the SM fermion fields, tell us that the Lie algebra of the SM gauge group is $\mathfrak{su}(3)\oplus \mathfrak{su}(2) \oplus \mathfrak{u}(1)$. The four gauge groups 
\begin{equation} \label{SM gauge group}
G=\frac{G_{\text{SM}}}{\Gamma_n}, \qquad G_{\text{SM}}=SU(3)\times SU(2) \times U(1), \quad \Gamma_n \cong \Z/n,\quad n\in \left\{1,2,3,6\right\},
\end{equation}
all share this Lie algebra and have representations corresponding to the SM fermions,\footnote{The embeddings of the discrete subgroups $\Gamma_n$ in $G_{\text{SM}}$ are given by Eq. (\ref{eq:xi}).} and any one of these may be the gauge group of the SM.\footnote{Indeed, even this is far from an exhaustive list. What is true is that the connected component of the SM gauge group $G$ is one of the four possibilities given in Eq. (\ref{SM gauge group}). } Thus, in addition to the various deficiencies in the SM that necessitate its extension, there is also an ambiguity in the SM. The potential physical distinctions between the four options in Eq. (\ref{SM gauge group}) were studied recently in Ref.~\cite{Tong:2017oea}, and amount to different periodicities of the $\theta$ angle associated with the hypercharge factor, and different spectra of Wilson lines in the theory. Perhaps unsurprisingly, all of these effects have a topological flavour.
 
Another possible distinction, which is also topological in origin but which was not discussed in Ref.~\cite{Tong:2017oea}, is that some of these options might not in fact be consistent after closer inspection, in the sense that they might suffer from anomalies. Of course, since the four groups in Eq. (\ref{SM gauge group}) share the same Lie algebra the conditions for local anomaly cancellation will be the same, and thus all these SMs are free of local anomalies, as is well known. However, this does not rule out the possibility of more subtle global anomalies in the SMs associated with the topology of the gauge group, analogous to (but much more general in scope than) the $SU(2)$ anomaly discovered by Witten~\cite{Witten:1982fp}, which might render some of the SM variants recorded in Eq. (\ref{SM gauge group}) inconsistent. Our first goal in this paper is to investigate the possible global anomalies for each choice of discrete quotient in (\ref{SM gauge group}), for arbitrary fermion content. 

To do so, we exploit the relation that arises in the absence of local gauge anomalies between the potential anomaly of the partition function (which arises in the phase) of a chiral gauge theory and the exponentiated $\eta$-invariant~\cite{atiyah_patodi_singer_1975} 
(which is a regularized sum of positive eigenvalues minus negative eigenvalues)
associated to an extension of the Dirac operator to a five-manifold that bounds spacetime.
This relation, which was first suggested in Ref.~\cite{Witten:1985xe}, follows from a set of mathematical results due to Dai and Freed \cite{Dai:1994kq}, which we  briefly review in \S \ref{sec:bordism} (for a more detailed discussion, see \cite{Witten:2015aba,Witten:2016cio,Witten:2019bou}).
To wit, one may show (via a vast generalisation of Witten's original `mapping torus' argument~\cite{Witten:1982fp}) that if $\exp 2\pi i \eta=1$
on all closed five-manifolds that are equipped with a spin structure and a map to $BG$,\footnote{To see why $BG$ is relevant, note that a gauge field is defined by a connection on a principal $G$-bundle over a spacetime manifold $\Sigma$, and every such bundle corresponds to a map
$\Sigma\rightarrow BG$; for global anomalies, the connection plays no role, and we have a one-to-one correspondence between $G$-bundles (without connection) and homotopy classes of maps $\Sigma\rightarrow BG$.} then there will be no anomalies on spacetimes which bound (in the sense that the requisite spin and gauge structures can be extended).
Since $\exp 2\pi i \eta$ is invariant under bordism in the case that local anomalies vanish, this is guaranteed to be the case when the group $\Omega^{\text{Spin}}_5(BG)$ (of equivalence classes under bordism of five-manifolds equipped with a spin structure and a map to $BG$) vanishes.\footnote{In fact, there are reasons to believe that the vanishing of $\Omega^{\text{Spin}}_5(BG)$ is sufficient for the vanishing of global anomalies not only on spacetimes that bound, but also on those that do not -- we discuss this at the end of \S \ref{sec:bordism}.}

In this paper we begin by applying this criterion for global anomaly cancellation to the four versions of the SM given by Eq. (\ref{SM gauge group}). 
The computations we report in this paper build upon those of Ref.~\cite{Garcia-Etxebarria:2018ajm}, which used the Atiyah-Hirzebruch spectral sequence to compute $\Omega^{\text{Spin}}_{d\leq 5}(BG)$ for a number of simple gauge groups $G$ including $SU(n)$, $PSU(n)$, $USp(2k)$, and $SO(n)$, as well as for $U(1)$. From there it was argued in Ref.~\cite{Garcia-Etxebarria:2018ajm} that there are no global anomalies in the SMs, by exploiting the (perhaps fortuitous) fact that the particular fermion content of the SM can be embedded in an anomaly-free grand unified theory (GUT) with $G=SU(5)$ (which breaks down to $G_{\text{SM}}/\Gamma_6$ as we go below the GUT scale). Alternative derivations of this result can be found in Refs.~\cite{Freed:2006mx,Wang:2018cai}.
It turns out that this guarantees that
all 4 versions of the SM in Eq. (\ref{SM gauge group}) are anomaly-free for the SM fermion content, or any other fermion representations that form representations of $SU(5)$.

We analyse the global anomalies in theories with one of the SM gauge groups by computing each $\Omega^{\text{Spin}}_5(BG)$ for the four gauge groups listed in Eq. (\ref{SM gauge group}) directly. At least in 3 out of the 4 cases (those in which $n\in\{1,2,3\}$), we can do this by first noting that the gauge group can be written as a 
product 
(for example, $G_{\text{SM}}/\Gamma_2 \cong U(2) \times SU(3)$). Next, we extend the methods of Ref.~\cite{Garcia-Etxebarria:2018ajm} to treat gauge groups which are products, by exploiting the fact that $B(G\times H)=BG \times BH$,\footnote{Similar ideas were used in the context of classifying higher-symmetry-protected topological phases~\cite{Wan:2018bns}.} and using a K\"unneth formula in (co)homology. The 4th case, in which $G=G_{\text{SM}}/\Gamma_6$, succumbs to a slightly more sophisticated attack, which we describe in \S \ref{sec: SM Z6}.

Our results for the four possible connected SM gauge groups can be applied, unlike those of Ref.~\cite{Garcia-Etxebarria:2018ajm}, to any BSM theories with one of the SM gauge groups but with different fermion content (that do not necessarily fit inside any GUT with a simple gauge group). While one might have expected, given the much more general nature of the anomaly cancellation condition imposed, more constraints to appear beyond those required to cancel the familiar $SU(2)$ global anomaly discovered by Witten, one finds that in fact that the opposite happens: in some cases there are actually fewer constraints, due to a subtle interplay between global and local anomalies, which we describe in \S \ref{sec:interplay}. This is related to the more mundane fact that for the gauge groups featuring quotients by $\Gamma_{n\neq 1}$ there are non-trivial constraints on the hypercharges of fermions depending on their representation. We give these constraints in \S \ref{sec:hypercharge}. 

We then turn our attention to global anomalies in a number of well-motivated BSM theories, which we analyse using the same bordism-based criteria. We demonstrate our methods in a wide variety of BSM examples, in the hope that readers can adapt the methods to analyse their own favourite models.
In particular, we consider theories in which the SM gauge group is extended by products with arbitrary $U(1)$ factors, as well as a number of GUTs including Pati-Salam models and trinification models. %, and a five-dimensional theory based on $SO(18)$. 

One might {\em a priori} expect all bets to be off when one goes beyond the SM, and that the possibility of $\Omega^{\text{Spin}}_5(BG)$ being non-trivial might provide a variety of extra constraints on the fermion content of BSM models for the cancellation of new global anomalies. Interestingly, we will find that this is largely not the case. In all the four-dimensional examples we considered, we find that $\Omega^{\text{Spin}}_5(BG)$ detects no new anomalies beyond the $\Z/2$-valued anomalies associated with $SU(2)$ (or more generally $Sp(r)$) factors in the gauge group. While we essentially arrive at a large collection of ‘null results’, we hope that the absence of any potential new anomalies in all of our examples will at least provide some assurance for the more conscientious BSM model-builders, who worry that their models might suffer from secret global anomalies.

We remark that in spacetime dimensions lower (or indeed higher) than four there are, however, potentially lots of new anomalies in theories with these gauge groups. We catalogue the relevant bordism groups in lower dimensions for the gauge groups we consider alongside the results of importance to the (B)SM case, in case they might be of interest to others (for example, in the condensed matter community). For ease of reference, all our bordism group results are collated across Tables \ref{summary bordism}, \ref{summary bordism generalised}, and \ref{summary bordism BSM}.

% In higher dimensions, we find that there are potential global anomalies in five-dimensional theories with the gauge group $SO(18)$, by computing that $\Omega_6^{\text{Spin}}(BSO(18))$ does not vanish. Thus, any GUT based on such a gauge group (such as that proposed in Ref.~\cite{Reig:2017nrz}) must be carefully checked for global anomalies.

The outline of the rest of this paper is as follows. In \S \ref{sec:bordism} we review the so-called `Dai--Freed theorem', and the arguments that underlie the bordism-based criterion for global anomalies that we  use. In \S \ref{sec:methodology} we review the algebraic machinery of spectral sequences which we  use to compute the bordism groups of interest to us. We then summarise and interpret our computations pertaining to global anomalies in the SMs in \S \ref{sec:SMs}. In \S \ref{sec:general}, we generalise the SM results to a 2-parameter family of theories that contains the SM, with gauge group $SU(N)\times Sp(M)\times U(1)$ for $N,\ M\in\Z$. We present the details of our computations for BSM theories in \S \ref{sec:BSM}. Finally, we find that there are no global anomalies in a BSM theory in which the SM fermions are defined using a spin$_c$ structure, allowing also for arbitrary additional fermion content, by showing that $\Omega_5^{\text{Spin}_c}(BG)=0$ for each choice of $G$ in Eq. \eqref{SM gauge group}. Such a theory can be defined on all orientable four-manifolds (not only those that are spin), but requires an additional $U(1)$ symmetry be gauged such as $B-L$.

\bigskip

{\em Note added}: Ref.~\cite{Wan:2019fxh}, which has subsequently appeared, confirms some of the bordism group calculations in this paper using the Adams spectral sequence.

\section{Bordism and global anomalies} \label{sec:bordism}

Both the local gauge anomalies first discovered by Adler, Bell, and Jackiw (ABJ)~\cite{PhysRev.177.2426,Bell:1969ts} and the global anomalies first discovered by Witten~\cite{Witten:1982fp} may arise in chiral gauge theories due to subtleties in defining the Dirac operator. To see how, and to motivate the more general bordism-based criterion for anomaly cancellation that we employ, 
it is helpful to first review some basic facts about chiral fermions, for which we largely follow the discussion in Ref.~\cite{Witten:2015aba}. Other helpful references for this discussion are Refs.~\cite{Witten:2016cio,Witten:2019bou,Yonekura:2016wuc} (written with physicists in mind) and the original mathematical paper by Dai and Freed on which much of the discussion rests~\cite{Dai:1994kq}.

Firstly, we recall that defining a chiral gauge theory requires that any spacetime manifold be equipped with certain geometric structures. The important structures for our purposes are
\begin{itemize}
\item A form of spin structure to define fermions,
\item A principal $G$-bundle to define gauge fields,
\item A Dirac operator which couples fermions to gauge fields, whose determinant is a well-defined function on the background data if the theory is to be non-anomalous.
\end{itemize}
We work in four spacetime dimensions from the beginning, since that is the case of relevance to the particle physics applications we are interested in; however, all the material we review in this Section generalises straightforwardly to other numbers of dimensions. We  always assume spacetime is euclideanised, and thus consider spacetime to be a smooth, compact, four-manifold $\Sigma$. At times it will be helpful to suppose $\Sigma$ is equipped with a (riemannian) metric, but this shall not be especially important to our arguments.

In most of this paper, we  assume that spacetime is orientable and that fermions are defined using an honest spin structure. It is possible, however, that fermions may be defined on an orientable spacetime using `weaker' structures if there are gauge symmetries present, as is typically the case in particle physics. For example, the presence of a $U(1)$ gauge symmetry allows one to define fermions using only a spin$_c$ structure; note that all orientable four-manifolds are spin$_c$, but not all orientable four-manifolds are spin. In \S \ref{sec:spin}, we  consider this possibility. In the presence of a larger gauge symmetry, such as $SU(2)$, one could get away with only a spin-$SU(2)$ structure to define fermions~\cite{Wang:2018qoy}, and so on.\footnote{A new kind of global anomaly has been recently discovered by Wang, Wen, and Witten~\cite{Wang:2018qoy} for an $SU(2)$ gauge theory formulated on all manifolds admitting such a spin-$SU(2)$ structure. They show that such a theory is anomalous if there is an odd number of fermion multiplets in spin $4r+3/2$ representations of $SU(2)$ (where $r\in\Z$). Of course, the more familiar $SU(2)$ global anomaly arises when the theory is defined on all spin manifolds, in which case there is an anomaly when $n_L-n_R=1$ mod $2$, where $n_L$ ($n_R$) is the number of left-handed (right-handed) $SU(2)$ doublets~\cite{Witten:1982fp}.
} 
In a time-reversal symmetric theory,\footnote{We note that the SM is not time-reversal symmetric, since $CP$ is explicitly broken by the phases appearing in the CKM and PMNS matrices, and in theory also by a non-zero QCD $\theta$ angle. Thus, in this paper we  only consider theories with one of the SM gauge groups to be defined on orientable spacetimes.} one could consider defining the theory also on {\em un}orientable spacetimes, in which case a form of pin structure could be used to define fermions. We describe how fermions can be defined using these various `spin structures' in Appendix \ref{app:spin} for reference; we also invite the reader to consult Appendix A of Ref.~\cite{Witten:2015aba}. Throughout the main body of this paper, however, we  assume that spacetime is orientable and equipped with a spin structure.

Defining gauge fields for some gauge group $G$ requires the existence of a principal $G$-bundle over $\Sigma$. As we wrote before, the  classifying space $BG$ of the Lie group $G$ has the property that the homotopy classes of maps from a space $X$ to $BG$ are in one-to-one correspondence with the set of (isomorphism classes of) principal $G$ bundles over $M$.\footnote{The classifying space $BG$ is the quotient of a weakly contractible space $EG$ by a proper free action of $G$. Any principal $G$-bundle over $M$ is the pullback bundle $f^\ast EG$ along a map $f:M\rightarrow BG$.} Thus, we consider orientable spacetimes $\Sigma$ equipped with a map $f:\Sigma \rightarrow BG$, in addition to a spin structure. We  moreover insist that a gauge theory be defined on {\em all} manifolds admitting these structures, leading to a very broad notion of whether there is an `anomaly' in the theory.
Ultimately, these requirements are necessary to guarantee that the theory be consistent with locality.

\subsection{Fermionic partition functions}

One may define fermions and gauge fields on four-manifolds equipped with the given geometric structures. In a renormalisable four-dimensional chiral gauge theory, one couples the two via the lagrangian $\bar{\psi} i \slashed{D} \psi$, where $i\slashed{D}$ is an hermitian Dirac operator. We are now in a position to see how both the local and global anomalies can emerge in such a gauge theory.

The heart of the trouble in both kinds of anomaly lies in performing the functional integration over fermions.
The result is a partition function $Z_\psi[A]$, which we  consider to be a function of the background gauge field and also any other background fields or data such as a metric on spacetime.\footnote{Sometimes, we  use `$A$' to denote the background gauge field, while at others time we  use `$A$' to collectively denote all the background fields/data. Which of the two meanings is implied in a given instance ought to be clear from the context. } Formally, $Z_\psi[A]$ is defined to be
\begin{equation} \label{psi partition function}
Z_\psi[A] \equiv \int \mathcal{D}\psi \mathcal{D} \bar{\psi} e^{-\int d^4x\ \bar{\psi} i \slashed{D} \psi} = \text{det}\ i\slashed{D},
\end{equation}
the determinant of the Dirac operator,\footnote{More generally, $Z_\psi[A]$ will be the Pfaffian of the Dirac operator. We  essentially ignore this subtlety for the purpose of this discussion, by assuming fermions to be complex or pseudo-real.} assumed to be appropriately regularized. The partition function $Z_\psi[A]$ of a non-anomalous quantum field theory is a kosher $\mathbb{C}$-valued function on the space of background data. For the case of coupling to background gauge fields, this means that $Z_\psi[A]$ must be a well-defined function on the space of connections on principal $G$-bundles modulo gauge transformations. 

If this is not the case, $G$-invariance is anomalous, and since it is a gauge symmetry, the theory is not well-defined. This viewpoint sets the traditional ideas of local and global gauge anomalies in a more general context: in the case of a local anomaly, one has that $Z_\psi[A] \neq Z_\psi[A^g]$ even for a gauge transformation $A\rightarrow A^g$ with $g$ infinitesimally close to the identity; for the original $SU(2)$ global anomaly~\cite{Witten:1982fp}, one finds $Z_\psi[A]=-Z_\psi[A^U]$ where the group element $U(x)$ corresponds to a gauge transformation in the non-trivial class of $\pi_4(SU(2))$. The partition `function' 
of an anomalous theory is thus at best a section of a complex line bundle over the space of background data, called the determinant line bundle. 
Moreover, the modulus $|Z_\psi|$ of the partition function cannot suffer from anomalies,\footnote{To see why, note that for any set of chiral fermions $\psi$, one can define a conjugate set $\widetilde\psi$ that transforms as the complex conjugate of $\psi$ under all symmetries, and with an action that is the complex conjugate of the action for $\psi$. Thus, the functional integration over $\widetilde\psi$ yields precisely $\bar{Z}_\psi$, the complex conjugate of (\ref{psi partition function}). Hence, for the combined system, the partition functon is $Z_\psi \bar{Z}_\psi = |Z_\psi|^2$. But given the complex conjugate set of fermions one can always write down mass terms for the set of fermions $\psi$, for which a Pauli-Villars regulator (which respects the symmetries of the lagrangian) is always available. Hence $|Z_\psi|^2$, and thus $|Z_\psi|$, cannot suffer from any anomalies.} and the anomaly must come purely from the  phase of $Z_\psi$.

With this realisation, one might first try to simply define the fermionic partition function to be equal to its modulus, and so construct an anomaly-free theory by {\em fiat}. But the modulus $|Z_\psi|$ on its own is not a smooth function of the background data $A$, just as $|w|$ is not a smooth function of the real or imaginary parts of a complex number $w$. The partition function must, however, depend smoothly on the background data, which includes gauge fields and metrics, otherwise correlation functions involving the stress-energy tensor and/or currents coupled to the gauge field would not be well-defined. Thus, one cannot evade anomalies in such a way, and one must instead consider carefully when $Z_\psi$ is well-defined, and when it is not. 

A set of mathematical results due to Dai and Freed~\cite{Dai:1994kq} allow one to construct a candidate partition function, which is necessarily smooth on the space of background data, with which to properly analyse anomalies. For brevity's sake, we  refer collectively to these results as the {\em Dai--Freed theorem}. For an account written with physicists in mind, see Ref.~\cite{Yonekura:2016wuc}.
\begin{figure}[h]
\includegraphics[width=0.35\textwidth]{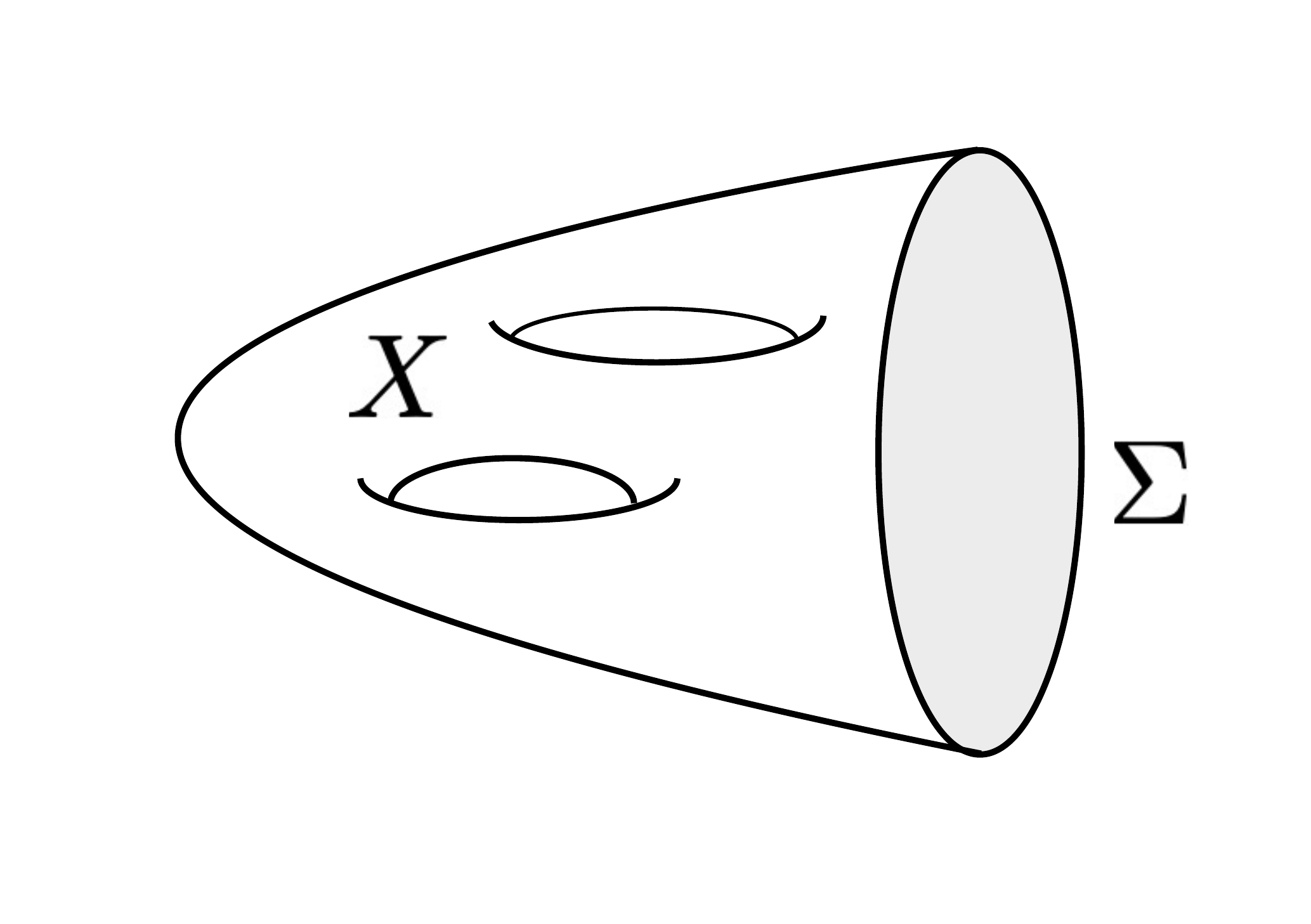}
\centering
  \caption{The results of Dai and Freed give a prescription for writing down a fermionic partition function $Z_\psi$ when spacetime $\Sigma$ is the boundary of a five-manifold $X$.}
  \label{fig:cigar}
\end{figure}

The Dai--Freed theorem implies that a putative partition function $Z_\psi[A]$ that is smooth in $A$ can always be defined when the four-dimensional spacetime $\Sigma$ is the boundary of a five-manifold $X$, {\em viz.} $\Sigma=\partial X$ (as depicted in Fig.~\ref{fig:cigar}), to which the theory (and thus the spin structure and map to $BG$) must be extended. The five-manifold $X$ must approach a `cylinder' $(-\tau_0,0]\times \Sigma$ near the boundary $\Sigma$, where the local coordinate $\tau\in(-\tau_0,0]$ parametrises the fifth dimension. Moreover, the Dirac operator is extended to define a five-dimensional Dirac operator on $X$ which we denote by $i\slashed{D}_{X}$, which near the boundary takes the form $i\slashed{D}_{X} = i\gamma^5(\partial_\tau + i\slashed{D})$, where $i\slashed{D}$ is the original Dirac operator on $\Sigma$.\footnote{Special boundary conditions must be chosen to ensure that the operator $i\slashed{D}_{X}$ is hermitian throughout $X$. These are often referred to as `(generalised) APS boundary conditions', and we will not discuss them further, but rather refer the reader to {\em e.g.} Refs.~\cite{Witten:2015aba,Yonekura:2016wuc}, in addition to the original papers of Atiyah, Patodi, and Singer~\cite{atiyah_patodi_singer_1975,atiyah_patodi_singer_1975ii,atiyah_patodi_singer_1976}.}

Schematically, the Dai--Freed definition of the putative partition function is then
\begin{equation} \label{dai freed}
Z_\psi[A] = |Z_\psi| \exp \left(-2\pi i \int_X I^0(F)\right) \exp \left(-2\pi i \eta_X \right),
\end{equation}
where we have split the phase into two distinct contributions, which we will define shortly. Importantly, Dai and Freed showed that this construction varies smoothly with the background data. 

The two contributions to the phase, as separated out in Eq. (\ref{dai freed}), correspond loosely to local and global anomalies. The first contribution to the phase of (\ref{dai freed}) is easier to understand. It is the integral of the anomaly polynomial $I^0(F)$ over the extended five-manifold $X$, which is a polynomial in the curvature $F$ of the connection $A$ defined such that 
\begin{equation} \label{ABJ phase}
dI^0(F)= \hat{A}(R)\text{~tr~}\exp \left(\frac{iF}{2\pi}\right) \bigg\rvert_6,
\end{equation}
where $\hat{A}(R)$ is the $\hat{A}$ genus (sometimes referred to as the `Dirac genus'), with $R$ the Riemann tensor. The bar and subscript `6' indicates that one should take only the six-form terms on the right-hand-side. This contribution to the phase is not necessarily invariant even under infinitesimal gauge transformations. Rather, its variation can be computed using Eq. (\ref{ABJ phase}), and requiring that this variation vanish after being integrated reproduces the familiar formulae for the cancellation of local anomalies (including gravitational and mixed gauge-gravitational anomalies). This type of anomaly is sometimes referred to as the perturbative anomaly, because one can derive it perturbatively by expanding the path integral around the zero background fields in flat spacetime.

\begin{figure}[h]
\centering
\begin{subfigure}[b]{0.75\textwidth}
  \centering
   \includegraphics[width=0.6\textwidth]{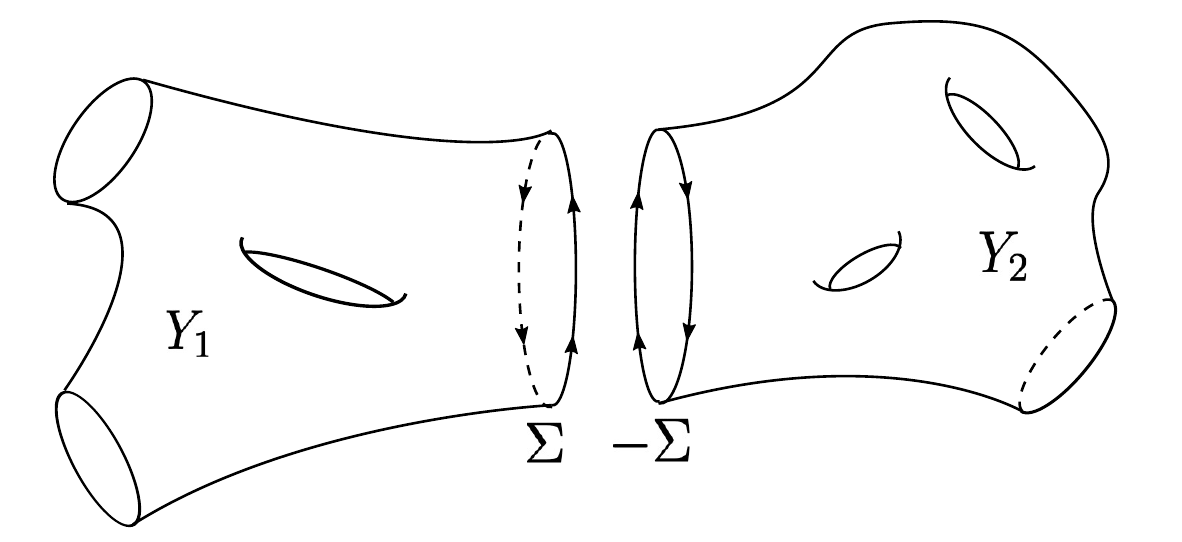}
   %\caption{}
   %\label{fig:Ng1} 
 \end{subfigure}

\begin{subfigure}[b]{0.75\textwidth}
  \centering
   \includegraphics[width=0.6\textwidth]{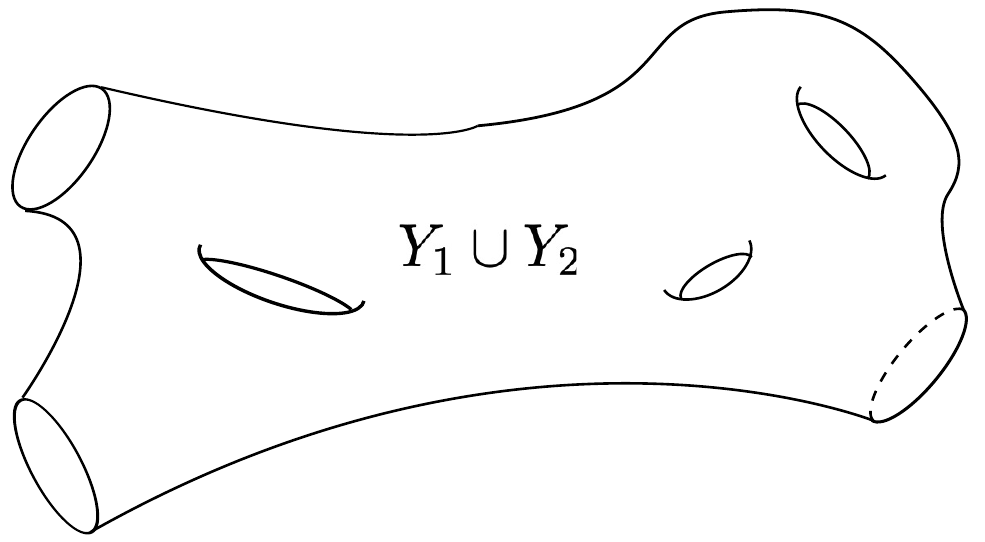}
   %\caption{}
   %\label{fig:Ng2}
\end{subfigure}
\caption{Gluing of two manifolds $Y_1$ and $Y_2$ with a shared boundary component $\Sigma$, under which the exponentiated $\eta$-invariant factorizes.} \label{fig:glue}
\end{figure}

The second contribution comes from the fermions on $X$, which one can think of as a kind of regulator for the system on $\Sigma$. The $\eta$-invariant is defined as the following sum over eigenvalues $\lambda$ of the Dirac operator $i\slashed{D}_X$
\begin{equation} \label{eta}
\eta_X = \frac{1}{2}\left(  \sum_{\lambda\neq 0} \text{sign}(\lambda) + \text{Dim} \ker (i\slashed{D}_X) \right),
\end{equation}
which must of course be regularized.\footnote{For example, in the original APS index theorem the sum over eigenvalues was regularized by replacing  $\sum_{\lambda\neq 0} \text{sign}(\lambda)$ with $\text{lim}_{s\rightarrow 0} \sum_{\lambda\neq 0} \text{sign}(\lambda) |\lambda|^{-s}$, which converges for large Re $s$, from which one can analytically continue to $s=0$ without encountering any poles.}
This $\eta$-invariant was introduced by Atiyah, Patodi, and Singer (APS) in their generalisation of the Atiyah--Singer index theorem to manifolds with boundary~\cite{atiyah_patodi_singer_1975,atiyah_patodi_singer_1975ii,atiyah_patodi_singer_1976}. It shall be useful in what follows to recall that the $\eta$-invariant possesses an important `gluing' property, as follows: if two manifolds with boundary $Y_1$ and $Y_2$ are glued along a common boundary to give a manifold $Y_1 \cup Y_2$, then the exponentiated $\eta$-invariant  factorizes, {\em i.e.}
\begin{equation} \label{gluing}
\exp \left( 2\pi i \eta_{Y_1\cup Y_2}\right)=\exp \left( 2\pi i \eta_{Y_1}\right)\ \exp \left( 2\pi i \eta_{Y_2}\right),
\end{equation}
as illustrated in Fig.~\ref{fig:glue}.

\subsection{Global anomalies and the \texorpdfstring{$\eta$}{eta}-invariant}

In order for (\ref{dai freed}) to describe an intrinsically four-dimensional theory on $\Sigma$, 
this putative definition for the fermionic partition function must be independent of the choice of five-manifold $X$ and the extension to $X$ of whatever structures are necessary to define the theory on $\Sigma$. Any dependence on $X$ invariably leads to ambiguities and inconsistencies with locality and/or smoothness in the four-dimensional theory. Such inconsistencies are precisely what we call “anomalies”.

It is worth mentioning here that, if the condition for anomaly cancellation is not satisfied, we can no longer use Eq. \eqref{dai freed} as the partition function for our theory on the four-manifold $\Sigma$. Nonetheless, even in this context \eqref{dai freed} remains a useful equation, because it precisely quantifies the anomalies in terms of anomaly inflow. Heuristically speaking, it tells us that we can make sense of an anomalous fermionic theory if it arises as a boundary degree of freedom of another theory in one dimension higher, where the anomalies at the boundary are precisely cancelled by the contribution from the bulk. This is captured solely by the $\eta$-invariant when there is no local anomaly, justifying our moniker of `global' anomalies. This fact lies at the heart of our current understanding of topological insulators in condensed matter physics. 

\begin{figure}[h]
\centering
\begin{subfigure}[b]{0.75\textwidth}
  \centering
   \includegraphics[width=0.8\textwidth]{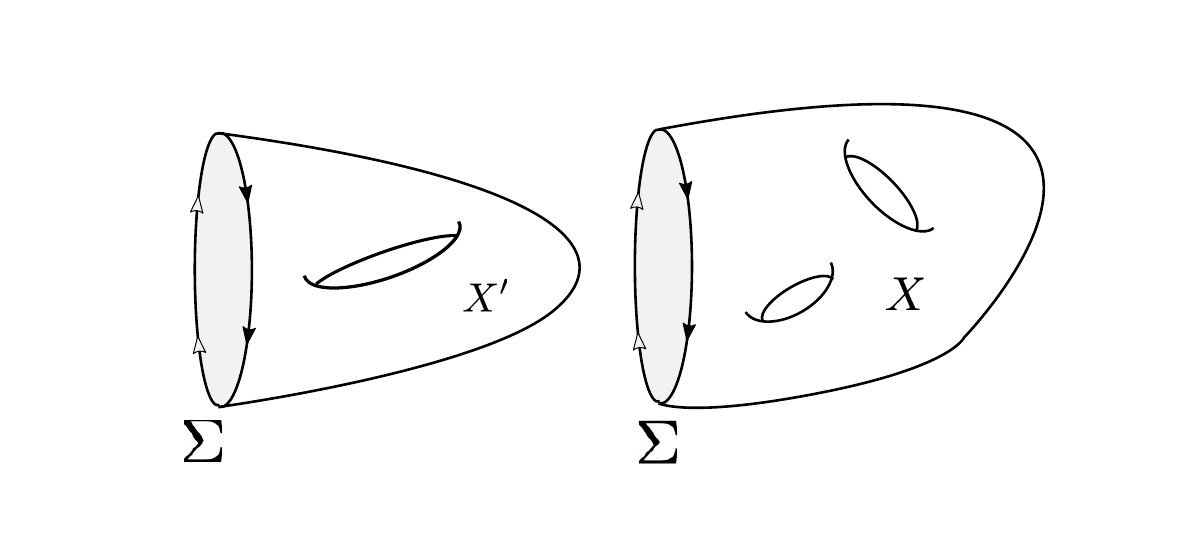}
   %\caption{}
   %\label{fig:Ng1} 
 \end{subfigure}

\begin{subfigure}[b]{0.75\textwidth}
  \centering
   \includegraphics[width=0.8\textwidth]{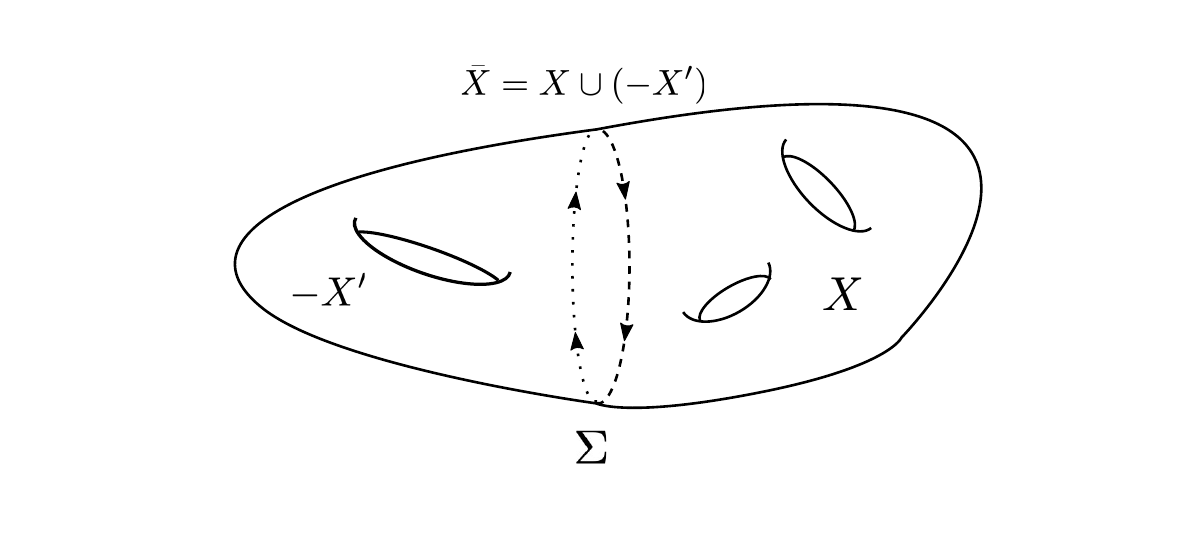}
   %\caption{}
   %\label{fig:Ng2}
\end{subfigure}
\caption{Gluing of two manifolds $X$ and $X^\prime$ with a shared boundary $\Sigma$ into a closed manifold $\bar{X} = X \cup (-X^\prime)$.} \label{fig:closed glue}
\end{figure}

Let us return to our search for a criterion for anomaly-freedom. The putative partition function \eqref{dai freed} is independent of the choice of five-manifold $X$ if and only if
\begin{equation}
    \label{eq: Y-indep}
    \exp\left(-2\pi i \eta_{\bar{X}}\right)= \exp\left(2\pi i \int_{\bar{X}} I^0(F)\right),
\end{equation}
for all {\em closed} five-manifolds $\bar{X}$. To see this, consider a duplicate of our fermionic theory on $\Sigma$ but extended to a different five-manifold $X^\prime$. Let $-X^\prime$ denote this five-manifold with its orientation reversed. It is then possible to glue the original system defined on $(X, \Sigma)$ to that on $(-X^\prime, -\Sigma)$ along the mutual four-boundary $\Sigma$. The result is a fermionic theory on a closed five-manifold $\bar{X} \equiv X \cup (-X^\prime)$, as illustrated in Fig.~\ref{fig:closed glue}.
Since the two systems have the same fermionic theory on $\Sigma$, the moduli of the path integrals cancel, and the path integral of the combined system is the pure phase
\begin{equation} \label{eq:anomaly theory}
    Z_{\bar{X}} = \frac{Z_X}{Z_{X^\prime}} = \exp \left(-2\pi i (\eta_X-\eta_{X^\prime})\right)\exp\left(2\pi i \left(\int_X-\int_{X^\prime}\right)I^0 (F)\right).
\end{equation}
Using the linearity property of integrals, together with the above gluing property for the $\eta$-invariant, we can rewrite the fermionic partition function on the closed five-manifold $\bar{X}$ as 
\begin{equation}
    Z_{\bar{X}} = \exp(-2\pi i\eta_{\bar{X}})\exp\left(-2\pi i \int_{\bar{X}} I^0(F)\right),
\nonumber
\end{equation}
which is trivial if and only if the condition \eqref{eq: Y-indep} is satisfied. The triviality of $Z_{\bar{X}}$ for any closed five-manifold $\bar{X}$ implies that $Z_{X} = Z_{X^\prime}$ for any pair of five-manifolds which share the same boundary theory  $\Sigma$. 

Thus, in the absence of local anomalies, {\em i.e.} when $I^0(F) = 0$, any residual global anomalies necessarily vanish, and the partition function describes an intrinsically four-dimensional theory, when $\exp\left(-2\pi i \eta_{\bar{X}}\right)=1$ for all closed five-manifolds $\bar{X}$ (that admit a spin structure and a map to $BG$). Witten's mapping torus argument~\cite{Witten:1982fp}, by which the original $SU(2)$ global anomaly was first detected (for a fixed spacetime $\Sigma=S^4$), is equivalent to insisting that $\exp\left(-2\pi i \eta_{\bar{X}}\right)=1$
on $\bar{X}=S^1\times S^4$. 

Moreover, when local anomalies cancel, such that $I^0(F) = 0$, it follows from the APS index theorem that $\exp(2\pi i\eta)$ is a bordism invariant.\footnote{This fact was first used in the physics literature to analyse global anomalies in  string theories~\cite{Witten:1985mj}.} By `bordism' we mean (unless explicitly stated otherwise) the equivalence relation on compact $p$-manifolds equipped with a spin structure and a map to $BG$ such that two manifolds are deemed equivalent if their disjoint union is the boundary of some compact $(p+1)$-manifold with the structures extended appropriately. By `bordism invariant', we mean a well-defined homomorphism on the equivalence classes under bordism (or just bordism classes), which form an abelian group $\Omega^{\text{Spin}}_p(BG)$.
This means that $\exp(2\pi i\eta)=1$ on any five-manifold that is null-bordant. 
Hence, when $I^0(F) = 0$ the $\eta$-invariant defines a homomorphism from the fifth spin bordism group to the phase of the partition function, or,
in other words
\begin{equation}
    \exp(2\pi i\eta)\in \text{Hom}\,\left(\Omega^{\Spin}_5(BG), U(1)\right).
\end{equation}
The group $\text{Hom}\,(\Omega^{\Spin}_5(BG), U(1))$ clearly vanishes if $\Omega^{\Spin}_5(BG)=0$. The vanishing 
of $\Omega^{\Spin}_5(BG)$ is in fact not only sufficient but also necessary for vanishing of $\text{Hom}\,(\Omega^{\Spin}_5(BG), U(1))$, at least when $\Omega^{\Spin}_5(BG)$ is a finitely generated abelian group (as is the case for all the examples we examine here),
which means it can be written as 
\begin{equation}
\Omega^{\Spin}_5(BG) \cong \Z^r \times \Z/p_1 \times \ldots \times \Z/p_m.
\end{equation}
To see that this is the case, note that for each summand there exist non-trivial maps to $U(1)$ -- for example, one can send $n\in \Z/p$ to $\exp(2\pi i n/p)$, or can send $k\in\Z$ to $\exp(\pi i k)$. Thus, as long as $\Omega^{\Spin}_5(BG) \neq 0$, the set of homomorphisms from the $5$th spin bordism group to $U(1)$ is non-empty.

The exponentiated $\eta$-invariant is necessarily trivial when $\Omega^{\Spin}_5(BG)$ vanishes.
Thus, if local anomalies cancel and if
\begin{equation}
\label{eq: triv bordism}
\Omega^{\Spin}_5(BG) = 0,
\end{equation}
then Eq. \eqref{eq: Y-indep} implies there is a well-defined fermionic partition function which is independent of the choice of five-manifold $X$, and thus defines a sensible local quantum field theory.

In summary, the following precise statement, which follows from the Dai--Freed theorem, forms the basis of what follows:
\begin{quote}
The path integral for a $d$-dimensional gauge theory with gauge group $G$ with arbitrary matter content can be consistently formulated on null-bordant spacetime manifolds of dimension $d$ using the Dai--Freed prescription if $I^0=0$ and $\Omega_{d+1}^{\text{Spin}}(BG) = 0$.
\end{quote}
Two caveats are warranted here. Firstly,
we still don't have a definition for spacetimes $\Sigma$ that are not null-bordant. Such spacetimes appear regardless of the gauge group,\footnote{Furthermore, in the presence of a non-abelian gauge symmetry, for example in the case $G=SU(3)$, there exist additional spacetime manifolds that do not bound spin five-manifolds (to which the map to $BG$ extends), generated by a manifold with instanton number one~\cite{Witten:2019bou}. }
being generated by a K3 surface~\cite{wild4}. 
In general, locality forces such spacetimes to appear in the theory, and so one needs a general prescription for the fermionic partition function evaluated on spacetimes in non-trivial bordism classes, which goes beyond the original Dai--Freed theorem. %\jc{(That said, such a prescription for the partition function on null-bordant spacetimes can be given, consistent with the principles of unitarity and locality, by assigning an arbitrary theta angle to each generator of $\Omega_{d}^{\text{Spin}}(\cdot)$; however, the appearance of arbitrary theta angles makes such a prescription not unique.)}

The second caveat is that, even if the Dai--Freed prescription cannot be made to work, it is still possible that some other suitable definition of the path integral might be found in cases where the condition (\ref{eq: triv bordism}) is violated. 

In fact, recent developments in the mathematical field of topological field theory give hints that these two caveats can safely be struck out. Those developments suggest that an anomalous theory should be viewed as a special case of a relative field theory \cite{Freed:2012bs}, namely a natural transformation between an extended field theory in one higher spacetime dimension (defined as a functor from some higher bordism category to some linear category) to the trivial extended field theory with the same dimension. Thus, part of the data of an anomalous field theory is a non-anomalous, non-trivial quantum field theory in one dimension higher. If there are no such theories, then there can be no anomalies. 

The putative theory in one dimension higher is, in many cases (but see Refs.~\cite{Freed:2012bs, Monnier:2014rua}), both topological and  invertible, meaning that it can be described by a classical topological action. It turns out that such actions can be classified by some Abelian group $A$ corresponding to some (generalized) differential cohomology theory. The group is characterised by an exact sequence of Abelian groups $B \to A \to C$, where $C$ corresponds here to the local anomaly and $B$ to the global anomaly. In the case of ordinary differential cohomology (in which we have not bordism classes of manifolds with spin, but rather homology classes corresponding to smooth singular simplices), the group $B$ is just the group $H^5(BG,U(1)) \cong \text{Hom} (H_5 (BG), U(1))$ and so it is tempting to conjecture that the corresponding group here is indeed $\text{Hom} (\Omega_5^{\text{Spin}} (BG), U(1))$. Moreover, in the ordinary differential cohomology case, the exact sequence $B \to A \to C$ extends to a short exact sequence $0 \to B \to A \to C \to 0$, so that $A=0$ iff. $B=C=0$. If the same is true here, then we have a complete characterisation of the anomaly cancellation conditions, whose global part is $\text{Hom} (\Omega_5^{\text{Spin}} (BG), U(1))=0$.

Indeed it is believed that~\cite{Freed:2016rqq,Witten:2019bou}, as long as the object $Z_ {\bar{X}}$ defined by (\ref{eq:anomaly theory}) equals one for all closed five-manifolds $\bar{X}$, a prescription for the partition function on non-nullbordant spacetimes can be given, that is consistent with the principles of unitarity and locality and free of anomalies, by assigning an arbitrary theta angle to each generator of $\Omega_{4}^{\text{Spin}}(BG)$. There is no quantum field theory principle that can be used to fix the arbitrary theta angles, which correspond to an element in $\text{Hom} (\Omega_4^{\text{Spin}} (BG), U(1))$, because any such element equals a partition function for an invertible topological field theory (in four dimensions) to which the theory may be consistently coupled. In the context of string theory these statements are well-known, with the assignment of theta angles sometimes referred to as ``setting the quantum integrand''~\cite{Witten:1996hc,Freed:2004yc}.

\section{Methodology} \label{sec:methodology}

It remains to explain how we actually compute a bordism group of the form $\Omega_5^{\text{Spin}}(BG)$, for a specific $G$. As is so often the case in algebraic topology, one is faced with a calculation that is seemingly impossible, no matter how simple the choice of $G$, but which turns out to be possible for almost any $G$, provided one knows enough tricks. The main tricks in the case at hand are the Atiyah-Hirzebruch spectral sequence~\cite{atiyah1961vector} (see Refs.~\cite{hatcher2004spectral,mccleary_2000} for introductions to spectral sequences) and the use of cohomology operations (see Ref.~\cite{Cartan1955}). 
We follow, essentially verbatim, the method set out in Ref.~\cite{Garcia-Etxebarria:2018ajm}, but we feel it might be helpful to readers to give a more pedestrian description, as follows.

\begin{figure}[h]
\includegraphics[width=1.05\textwidth]{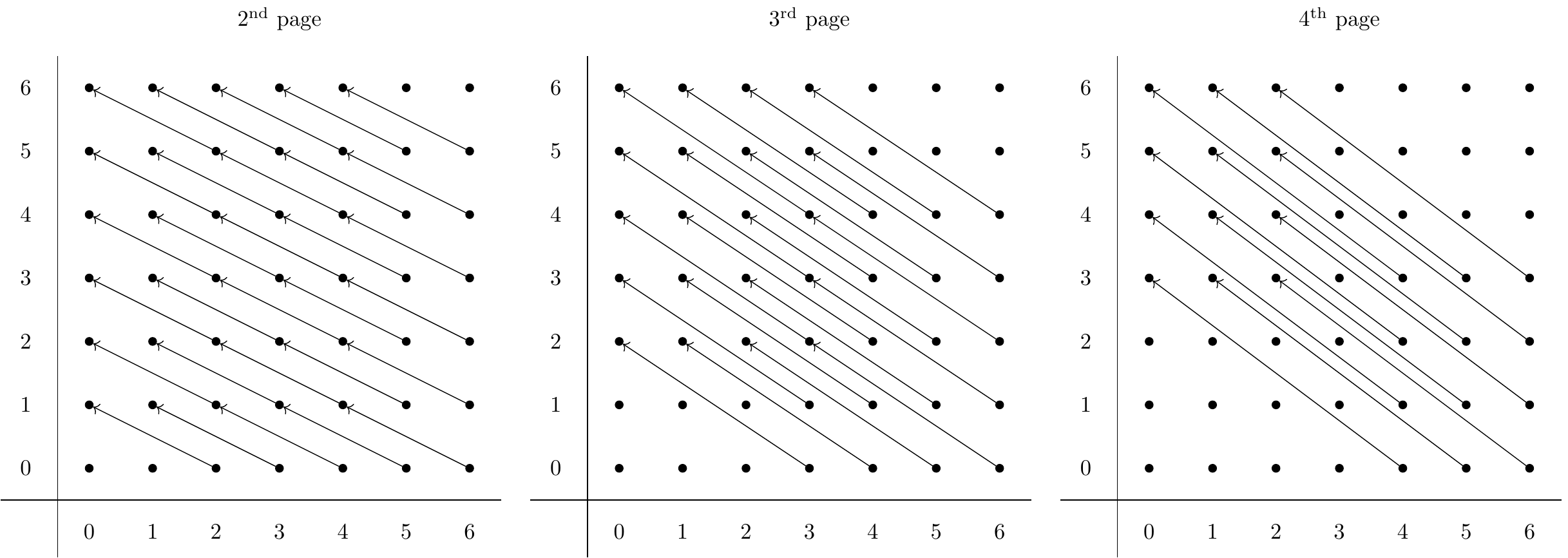}
\centering
  \caption{Schematic illustration of a spectral sequence
  }
  \label{fig:SS Sample}
\end{figure}

Spectral sequences are an important calculational tool in algebraic topology. So, what is a spectral sequence? In essence, a spectral sequence is a collection of abelian groups $E^r_{p,q}$ indexed by three non-negative integers $r$, $p$, and $q$, together with a collection of group homomorphisms between them. Perhaps more appealingly, one can picture a spectral sequence to be a `book' consisting of (infinitely) many pages, labelled by a `page number' $r$, with a two-dimensional array of abelian groups $E^r_{p,q}$ on each page. There are maps (called `boundary maps' or `differentials') between the groups within a given page of the form\footnote{Note that we are here describing the {\em homological} version of a spectral sequence, which shall also be the kind we employ in our bordism computations. There is an analogous {\em co}homological version, in which the boundary maps go in the opposite directions.
} 
\begin{equation} \label{differentials general}
d^r_{p,q}:E^r_{p,q}\rightarrow E^r_{p-r,q+r-1}, \qquad \text{such that} \qquad d^r_{p-r,q+r-1} \circ d^r_{p,q}=0,
\end{equation}
which endows the groups $E^r_{p,q}$ on the corresponding `diagonals' of a given page with the structure of a {\em chain complex}. The first few pages are illustrated schematically in Fig.~\ref{fig:SS Sample}. Moreover, one passes from one page to the next by `taking the homology' with respect to the differentials, specifically
\begin{equation}
E^{r+1}_{p,q} \cong \ker(d^r_{p,q})/ \Ima(d^r_{p+r,q-r+1}).
\end{equation}
As we keep `turning the pages' in this way, the abelian group appearing in any given $(p,q)$ position will eventually stabilise (because there are only a finite number of differentials going `in' and `out' for any $(p,q)$). It is conventional to refer to the `last page', after which all entries of the AHSS have stabilised, as $E^\infty_{p,q}$. Important topological information will be contained in this last page.

For example, the Serre spectral sequence can be used to compute the (co)homology groups of a topological space $X$ appearing as the total space in a 
fibration $F\rightarrow X \rightarrow B$, 
from the (co)homology of the two spaces $F$ and $B$, where we  take $B$ to be simply connected. 
For the Serre spectral sequence, we can in fact ignore the first page, and begin at the second page, whose entries are given by the peculiar formula $E^2_{p,q}=H_p(B;H_q(F;A))$; in words, the homology groups of the base space with coefficients valued in the homology groups of the fibre (for some coefficient group $A$). We then proceed to turn the pages using the differentials (\ref{differentials general}), until we get to the last page at which all the entries have stabilised. Then the $n$th homology group of the total space $X$ can be pieced together for each $n$, using $H_n(X;A)=\bigoplus_p E^\infty_{p,n-p}$, in others words, by taking the direct sum of all the groups on the $n$th diagonal of the last page of the Serre spectral sequence.\footnote{This is in fact a simplification, and only holds when the coefficient group $A$ is a field. Otherwise, a non-trivial group extension problem must be solved.} 

The Atiyah-Hirzebruch spectral sequence (AHSS) is a generalisation of the Serre spectral sequence just described, in which ordinary (co)homology is replaced by generalised (co)homology. The bordism groups $\Omega_5^{\text{Spin}}(BG)$ that we want to compute to classify global anomalies are examples of generalised homology groups, 
and so the AHSS provides an appropriate tool for our computation, if we can fit $BG$ into a useful fibration 
\begin{equation}
F\rightarrow BG \rightarrow B.
\end{equation}
Given such a fibration, the AHSS is then constructed in a similar fashion to the Serre spectral sequence. We begin at the second page, whose entries are now the homology groups
\begin{equation} \label{AHSS pg 2}
E^2_{p,q}=H_p(B;\Omega^{\text{Spin}}_q(F)).
\end{equation}
If the singular homology groups $H_p(B;\Z)$ are {\em free} ({\em i.e.} do not contain torsion) then this simplifies to
\begin{equation} \label{AHSS pg 2 free}
E^2_{p,q}=H_p(B;\Omega^{\text{Spin}}_q(F))=H_p(B;\Z) \otimes \Omega^{\text{Spin}}_q(F).
\end{equation}
If this is not the case, then the {\em universal coefficient theorem} (in homology) must be used to calculate (\ref{AHSS pg 2}). This second page comes equipped with differentials as specified in Eq. (\ref{differentials general}), and if the differentials are known we can turn to the next page. If we are able to continue turning pages until all the entries with $p+q=5$ are stabilised, then we can use these entries to extract $\Omega^{\text{Spin}}_5(BG)$. Analogous to the example of the Serre spectral sequence, it shall be the case in all the examples we consider that $\Omega^{\text{Spin}}_5(BG)$ shall simply be the direct sum of the entries $E^\infty_{p,q}$ with $p+q=5$.\footnote{While there is a straightforward condition telling us when this is the case for the Serre sequence - namely, when the coefficient group $A$ is a field - there is (as far as we are aware) no similarly straightforward condition pertaining to the AHSS and our bordism calculations. Rather, one must refer to the definition of the spectral sequence in terms of {\em filtrations} of the bordism groups we are trying to compute, using which the answer can often be extracted unambiguously from the last page. In particular, this was the case in all the examples we present in the sequel.  }

The simplest fibration involving $BG$, which we shall employ most frequently, is the trivial one in which $BG$ is fibred over itself, such that the fibre is a point which we denote by $\text{pt}$, {\em i.e.} we consider
\begin{equation} \label{point fibration}
\text{pt} \longrightarrow BG \longrightarrow BG.
\end{equation}
In this case, computing the elements (\ref{AHSS pg 2 free}) of the second page of the AHSS requires two ingredients: (i) the singular homology groups of the classifying space, $H_p(BG;\Z)$, and (ii) the bordism groups (preserving the spin structure) equipped with maps to a point; in other words, simply the equivalence classes (under bordism) of spin five-manifolds. Fortunately for us, these bordism groups are well known in low dimensions~\cite{anderson1966spin}:
\begin{equation}
\label{eq:bordism-table}
\def\arraystretch{1.5}
\arraycolsep=4pt
\begin{array}{c|ccccccccccc}
n & 0 & 1 & 2 & 3 & 4 & 5 & 6 & 7 & 8 & 9 & 10 \\
\hline
  \Omega^{\text{Spin}}_n(\text{pt}) & \Z & \Z /2 & \Z /2 & 0 & \Z & 0 & 0 & 0 & \Z^2 & (\Z /2)^2 & (\Z /2)^3\\
\end{array}
\end{equation}

The other ingredients we need are the homology groups of the classifying space of any gauge group $G$ we want to consider. As we have advertised above, we will consider many examples where $G$ is a product 
and our strategy here will be to build up the homology groups of such groups from the homology groups of their factors. We shall make frequent use of the fact that
\begin{equation} \label{eq:BG}
B(G\times H) = BG \times BH,
\end{equation}
which follows from the definition of the classifying space of a group (see, for example, Chapter 16, \S 5 of \cite{may1999concise}). Thence, we shall use the K\"unneth theorem to compute the homology of the product space $BG\times BH$ with coefficients in $\Z$.
In the absence of torsion,\footnote{If there is torsion, the correct statement of the K\"unneth theorem is that there is a short exact sequence
\begin{equation}
0 \rightarrow \bigoplus_{m+n=p} H_m(BG;\Z) \otimes H_n(BH;\Z) \rightarrow H_p(BG\times BH;\Z) \rightarrow \bigoplus_{m+n=p-1} \text{Tor}\left( {\scriptstyle H_m(BG;\Z),\;  H_n(BH;\Z)     }\right) \rightarrow 0,
\end{equation}
and that this sequence splits (although not canonically).}
 this is simply
\begin{equation}
H_p(BG\times BH;\Z) \cong \bigoplus_{m+n=p} H_m(BG;\Z) \otimes H_n(BH;\Z).
\end{equation}
The classifying spaces (and their homology rings) for some elementary groups are well-known; for example, $BU(1)=\mathbb{C}P^\infty$, with
\begin{equation}
H_p(BU(1)=\mathbb{C}P^\infty;\Z)=   \begin{cases}
    \Z \quad & \text{when } p = 0 \text{ mod } 2\, ,\\
    0 \quad & \text{otherwise}\, ,
  \end{cases}
\end{equation}
and $BSU(2)=\mathbb{H}P^\infty$, with 
\begin{equation}
H_p(BSU(2)=\mathbb{H}P^\infty;\Z)=   \begin{cases}
    \Z \quad & \text{when } p = 0 \text{ mod } 4\, ,\\
    0 \quad & \text{otherwise}\, .
  \end{cases}
\end{equation}
While the homology groups for these two examples are known in all degrees, it is often enough for our purposes to know the groups $H_p(BG;\Z)$ in sufficiently low dimensions; for instance, the result 
\begin{equation}
H_p(BSU(n);\Z)=\{\Z,0,0,0,\Z,\dots\}
\end{equation}
(for $n>1$) shall be useful for our consideration of gauge theories relevant to particle physics. 

Unfortunately for our purposes, results are usually quoted for {\em co}homology groups of classifying spaces, not least because of their starring role in the theory of characteristic classes. But one can obtain the homology groups using some universal coefficient theorem. 

\subsubsection*{Turning the pages}

We have now proposed how to obtain all the ingredients with which to write down the second page of the AHSS associated with the fibration (\ref{point fibration}); but we do not yet know how to turn to the next page of the AHSS, which requires knowledge of the differential maps introduced in Eq. (\ref{differentials general}). One thing we know for certain is that the differentials are {\em group homomorphisms}, and in many cases this shall turn out to be enough to deduce the image and/or kernel of many differentials unambiguously; for example, we make frequent use of the fact that $\text{Hom} (\Z/n,\Z) \cong 0$.
Similarly, for any pair of finite integers $n$ and $m$, we may use the fact that  $\text{Hom} (\Z/n,\Z/m) \cong \Z/\text{gcd}(n,m)$.

However, simple algebraic arguments like this will seldom be enough to determine all the differentials in the AHSS. Fortunately, we can make use of the fact that some of the differentials {\em on the second page} $E^2_{p,q}$ are known for the case of the spin bordism groups $\Omega_q^{\text{Spin}}$. In particular, we have that the differential
\begin{equation} \label{steenrod 0 to 1}
d^2_{p,0}:H_p(B;\Omega_0^{\text{Spin}})\rightarrow H_{p-2}(B;\Omega_1^{\text{Spin}})
\end{equation}
is the composition of the (homology) dual of the Steenrod square and followed by reduction modulo 2~\cite{teichner1992topological,teichner1993signature}, and that the differential 
\begin{equation} \label{steenrod 1 to 2}
d^2_{p,1}:H_p(B;\Omega_1^{\text{Spin}})\rightarrow H_{p-2}(B;\Omega_2^{\text{Spin}})
\end{equation}
is the dual of the Steenrod square~\cite{teichner1992topological,teichner1993signature}. The Steenrod square, Sq$^2$, is an operation on mod 2 {\em co}homology classes, Sq$^2:H^n(X;\Z/2) \rightarrow H^{n+2}(X;\Z/2)$, whose particular action on the generators of $H^n$ are known for the classifying spaces of Lie groups, thanks to Borel and Serre~\cite{borel1953groupes}. We will make regular use of their results in what follows. We note here for future reference that $\text{Sq}^2$ is an example of more general Steenrod squares, $\text{Sq}^k: H^{n}(X;\Z/2) \rightarrow H^{n+k}(X;\Z/2)$ which are operations on mod 2 cohomology rings satisfying the following properties
\begin{align}
  1)& \quad \text{Sq}^0 (x) = x,\nn\\
  2)& \quad \text{Sq}^k (x) = 0 \quad \text{if}\;\; k>\text{deg}(x),\nn\\
  3)& \quad \text{Sq}^{\text{deg}(x)} (x) = x\cup x ,\nn\\
  4)& \quad \text{Sq}^k (x\cup y) = \sum_{i+j=k} \text{Sq}^i(x) \cup \text{Sq}^j (y) \quad \text{(Cartan's formula)}\label{eq: Cartan}
\end{align}
Moreover, 
the Steenrod squares, being natural transformations of cohomology functors, have the property that they commute with the map $f^*: H^\bullet (Y;\Z/2)\rightarrow H^\bullet (X;\Z/2)$ induced  on cohomology by a map $f:X\rightarrow Y$. Thus we have $f^* \text{Sq}^k_Y = \text{Sq}^k_X f^*$.

By virtue of  this naturality, the Steenrod squares' action on $H^\bullet(BG_1\times BG_2;\Z/2)$, which we denote by $\text{Sq}^k_{\times}$ for clarity, are fully determined by their action on $H^\bullet(BG_1;\Z/2)$ and $H^\bullet(BG_2;\Z/2)$, denoted by $\text{Sq}^k_{1}$ and $\text{Sq}^k_{2}$. To see this, consider a projection $\pi_i: BG_1\times BG_2 \rightarrow BG_i$, with $i=1,2$. Let $c_i\in H^\bullet(BG_i;\Z/2)$ be a generator. By naturality we have $\text{Sq}^k_{\times} (\pi^* c_i) = \pi^* (\text{Sq}^k_{i} \,c_i)$.  But since $\pi_i^*c_i$ is naturally identified with $c_i$ through the K\"{u}nneth theorem for cohomology, this gets simplified to
  \begin{equation}
    \text{Sq}^k_{\times} c_i = \text{Sq}^k_{i} \,c_i.
    \label{eq: Sq with prod}
  \end{equation}
  With help from Cartan's formula \eqref{eq: Cartan}, the Steenrod squares' action on any generator of $H^\bullet(BG_1\times BG_2;\Z/2)$ can be subsequently worked out.

\section{Global anomalies in the Standard Model(s)} \label{sec:SMs}

Now that we have laid the groundwork and described the computational tools we  use to identify potential global anomalies, we are ready to report our computations. We begin with a gauge theory of indisputable importance to particle physics phenomenology, namely the Standard Model(s). Our results for the SM gauge groups are summarised in Table~\ref{summary bordism}.

The Standard Model (SM) of particle physics is a four-dimensional gauge theory, with gauge group
\begin{equation} \label{SM gauge group 2}
G=\frac{G_{\text{SM}}}{\Gamma_n}, \qquad G_{\text{SM}}=SU(3)\times SU(2) \times U(1), \quad \Gamma_n\cong \Z/n, \quad n\in\{1,2,3,6\}.
\end{equation}
Here, the $\Z/6$ quotient in the case of $\Gamma_6$ is generated by the element
\begin{equation} \label{eq:xi}
\xi = (\omega,\eta,e^{2\pi i/6})\in G_{\text{SM}},
\end{equation}
where $\omega$ is the generator of the $\Z/3$ centre of $SU(3)$ (with  $\omega^3=\textbf{1}\in SU(3)$), and $\eta$ is the generator of the $\Z/2$ centre of $SU(2)$ (with $\eta^2=\textbf{1}\in SU(2)$). 
The $\Gamma_3$ quotient in (\ref{SM gauge group 2}) is generated by $\xi^2$, and the $\Gamma_2$ quotient by $\xi^3$.
The fermion content of the SM consists of quarks and leptons, which are chiral fermions transforming in the following representations of $G$
$$
Q \sim (\mathbf{3}, \mathbf{2})_{1/6}, \quad
U^c \sim (\overline{\mathbf{3}}, \mathbf{1})_{-2/3}, \quad
D^c \sim (\overline{\mathbf{3}}, \mathbf{1})_{1/3}, \quad
% $$
% $$
L \sim (\mathbf{1}, \mathbf{2})_ {-1/2}, \quad
E^c \sim (\mathbf{1}, \mathbf{1})_{1},
$$
where here all the fields indicated are left-handed.

We compute the fifth bordism group (preserving spin structure) for all four groups listed in Eq. (\ref{SM gauge group 2}), and so identify potential global anomalies in these theories. Recall that in Refs.~\cite{Garcia-Etxebarria:2018ajm,Freed:2006mx}, it was argued that there are no global anomalies in the SM with any of these four gauge groups, by fitting all four possibilities inside an $SU(5)$ GUT which is easily shown to be anomaly-free (since the computation of the bordism group for $SU(n)$ is straightforward). What we shall prove is a more general result, since it shall apply to gauge theories with one of these four gauge groups, but with {\em arbitrary} fermion content. Thus, the results we find shall apply immediately to any BSM theories in which the gauge group is that of the SM, but in which there are additional chiral fermion fields.

\subsection{Hypercharge constraints} \label{sec:hypercharge}

Before we start computing bordism groups, it is important to point out that if we extend the SM by adding extra fermions, one must make sure that such fermions transform in {\em bona fide} representations of whichever gauge group from Eq. \eqref{SM gauge group 2} is being considered. 
In the cases where $G=G_{\text{SM}}/\Gamma_n$ with $n\in\{2,3,6\}$ there are constraints on the possible hypercharges fermions can take, depending on their representation under the $SU(3)\times SU(2)$ factor of $G_{\text{SM}}$. Since the derivations of these constraints involve a digression into representation theory, we relegate them to Appendix \ref{App: U(n) irreps}. In this Section we simply record what these constraints are -- specifically, see Eqns (\ref{eq: general U2 charge constraint}, \ref{eq: general U3 charge constraint}, \ref{eq: Z6 hypercharge constraint}). (Needless to say, the SM fermion representations satisfy these constraints.)

\subsubsection*{The $\Gamma_2$ quotient case}
Given the $\Z/2$ quotient in the case $G=G_{\text{SM}}/\Gamma_2$ is generated by $\xi^3$, where $\xi$ is given in Eq. (\ref{eq:xi}), 
we can write this particular quotient of the SM gauge group as 
\begin{equation} \label{eq: Z2 quotient GG}
\frac{G_{\text{SM}}}{\Gamma_2} =  SU(3)\times \frac{SU(2)\times U(1)}{\Z/2} \cong SU(3)\times U(2).
\end{equation}
In addition to its use in deriving the hypercharge constraints, writing the gauge group in this way ({\em i.e.} as a product) is crucial to our strategy for computing its bordism groups, in \S \ref{sec:Z2 SM quotient}.
Focussing on the $U(2) = \left( SU(2)\times U(1)\right)/(\Z/2)$ factor of $G$, a representation of  $U(2)$ corresponds to a representation of $SU(2)\times U(1)$, which in this subsection we denote by $(j,q)$ where $j$ denotes the isospin-$j$ representation of $SU(2)$ (which has dimension $2j+1$) and $q\in\Z$ is the integer-normalised $U(1)$ charge, with some restrictions imposed. 

To see how these constraints arise, let us first consider a field $\psi$ transforming in the representation $(\frac{1}{2},q)$, {\em i.e.} in the fundamental representation of $SU(2)$, since this is the simplest case. 
This means that  $\psi \mapsto \psi^\prime = \exp\left(i q \theta\right) \sigma\cdot \psi$
under the action of the $U(2)$ group element corresponding to $(\sigma,\exp i\theta)\in  SU(2)\times U(1)$.
For this to be a kosher representation of $U(2)$, one must identify the action of $\left(\textbf{1},\exp i\pi\right)$ and $\left(-\textbf{1},1 \right)$, which gives us the constraint
$\exp i q \pi = -1$.
  Therefore, any $SU(2)$ doublet must have hypercharge
  \begin{equation}\label{eq: Z2 hypercharge constraint}
  q=1 \text{~mod~} 2,
  \end{equation}
  {\em i.e.} an odd integer.\footnote{Similar restrictions on $U(1)$ charges appear in the context of defining fermions on manifolds that are not necessarily spin, by using the $U(1)$ gauge symmetry to define a spin$_c$ structure. In that context, such charge restrictions depend on the representations of fermions under the Lorentz group, and are thus referred to as `spin-charge relations'~\cite{Seiberg:2016rsg}. We  consider these spin-charge relations more in \S \ref{sec:spin}. }
   This is the case in the SM, where the doublet representations $Q$ and $L$ carry hypercharges $1$ and $-3$ respectively, using an integer normalisation in which the smallest charge (that belonging to $Q$) is set to one. 

If one wishes to add additional electroweak doublets, choosing the gauge group \eqref{eq: Z2 quotient GG}, one must ensure they too have odd hypercharges.
   
If one adds additional BSM fields transforming in larger representations of $SU(2)$, there are similar constraints on their hypercharges if they are to embed in representations of $U(2)$. To wit,  for a field transforming in the $(j,q)$ representation, the hypercharge must satisfy
     \begin{equation} \label{eq: general U2 charge constraint}
  q=2j \text{~mod~} 2.
  \end{equation}
  In other words, the charge must be even for all integer isospin representations (including, of course, any $SU(2)$ singlets), and odd for all half-integer isospin representations.
  For the proof of this general statement, we refer the reader to Appendix~\ref{App: U(n) irreps}.

\subsubsection*{The $\Gamma_3$ quotient case}

Given the $\Z/3$ quotient in the case $G=G_{\text{SM}}/\Gamma_3$ is generated by the element $\xi^2$, we can write this variant of the SM gauge group in the more useful form 
\begin{equation} \label{eq: Z3 quotient GG}
\frac{G_{\text{SM}}}{\Gamma_3}= \frac{SU(3)\times U(1)}{\Z/3}\times SU(2) \cong U(3)\times SU(2),
\end{equation}
In this case, we obtain hypercharge constraints on any fields transforming non-trivially under $SU(3)$, by requiring that they embed in representations of $U(3)$. 

Consider the simplest case of a field $\psi$ transforming in the fundamental triplet representation of $SU(3)$ ({\em a.k.a.} a quark) and with charge $q$ under $U(1)$. Under the action of $\exp(i q\theta)g\in U(3)$, for some $g\in SU(3)$, we have that $\psi \mapsto \psi^\prime = \exp(i q\theta)g\cdot \psi$. To be a {\em bona fide} representation of $U(3)$ means that $(\exp^{2\pi i/3},\mathbf{1}_3)$ and $(1,\omega=e^{2\pi i/3}\mathbf{1}_3)$ are identified in $SU(3)\times U(1)$, 
giving the constraint    $e^{2q \pi i/3} = e^{2\pi i/3}$. Hence, any colour triplet must have hypercharge 
\begin{equation}\label{eq: Z3 hypercharge constraint}
  q=1 \text{~mod~} 3.
\end{equation}  
The SM quark fields $Q$, $U$, and $D$ have hypercharges $+1$, $+4$, and $-2$ respectively, all of which are indeed equal to 1 mod 3.

One might consider adding fermions in other representations of $SU(3)$, and for each representation there is a corresponding hypercharge constraint.  Irreducible representations of $SU(3)$ correspond to Young diagrams with two rows, and so can be labelled by a pair integers $(\lambda_1,\lambda_2)$ corresponding to the number of boxes in each of the two rows, with $\lambda_1 \geq \lambda_2 \geq 0$. In Appendix \ref{App: U(n) irreps}, we prove that the hypercharge $q$ of a field transforming in the $(\lambda_1,\lambda_2)$ representation of $SU(3)$ must satisfy
\begin{equation} \label{eq: general U3 charge constraint}
q = (\lambda_1+\lambda_2) \text{~mod~} 3,
\end{equation}
if the gauge group is $U(3)\times SU(2)$. Note in particular that any colour singlets must have charge $q\in 3\Z$, as is the case for the SM leptons.

\subsubsection*{The $\Gamma_6$ quotient case}

Finally, we discuss the case with gauge group $G=G_{\text{SM}}/\Gamma_6$.
Consider a field in an arbitrary representation of this gauge group, corresponding to the $(\lambda_1,\lambda_2)$ representation of $SU(3)$, the isospin-$j$ representation of $SU(2)$, and with $U(1)$ charge $q$.
The hypercharge constraint is that
\begin{equation}
q = 2j \text{~mod~} 2 = (\lambda_1+\lambda_2) \text{~mod~} 3
\end{equation}
(see Appendix \ref{App: U(n) irreps}). For example, for a field with $j=1/2$ and $(\lambda_1,\lambda_2)=(1,0)$, {\em i.e.} corresponding to the bifundamental representation of $SU(3)\times SU(2)$, this constraint reduces to 
\begin{equation}\label{eq: Z6 hypercharge constraint}
  q=1 \text{~mod~} 6.
\end{equation}  
The only SM fermion transforming in the bifundamental representation of $SU(3)\times SU(2)$ is the left-handed quark doublet $Q$, and sure enough the charge of $Q$ is one.

\bigskip

Having established these constraints on the hypercharges of fermion fields for these four versions of the SM gauge group, we now  turn to our main concern, which is to compute the bordism groups of $BG$ for each of the four possible gauge groups $G$, which detect potential global anomalies theories with these gauge groups. We begin with the simplest case.

\subsection{\texorpdfstring{$\Omega_5^{\text{Spin}}(BG_{\text{SM}})$}{Bordism group of SM}} \label{sec:no quotient}

For the simplest case where $G=G_{\text{SM}}=SU(3)\times SU(2)\times U(1)$ with a regular spin structure, we  use the AHSS associated with the fibration (\ref{point fibration}) to compute the bordism groups $\Omega^{\text{Spin}}_{d\leq 5}(BG_{\text{SM}})$.

\begin{figure}[h]
\includegraphics[width=0.6\textwidth]{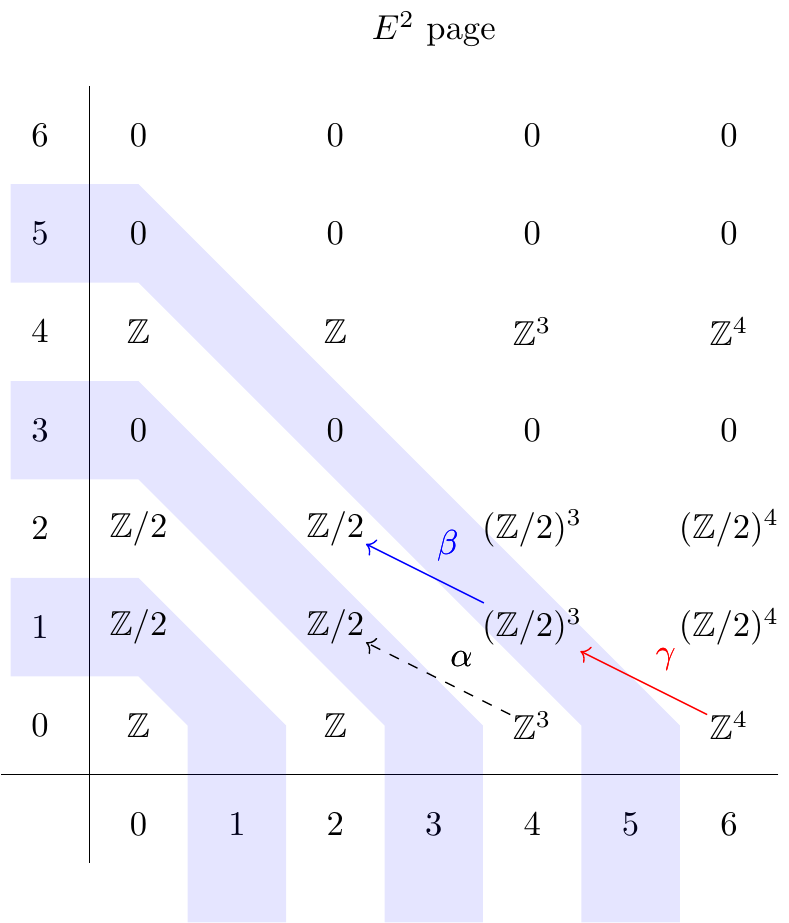}
\centering
  \caption{The $E^2$ page of the Atiyah-Hirzebruch spectral sequence for $G=G_{\text{SM}}$. We see that there is only a single entry relevant to the computation of $\Omega_5^{\text{Spin}}(BG_{\text{SM}})$, with a map ($\gamma$) going in and a map ($\beta$) going out.
  }
  \label{fig:AHSS SM}
\end{figure}

To begin, we have that
\begin{equation}
B \left[ SU(3)\times SU(2)\times U(1) \right] = BSU(3) \times BSU(2) \times BU(1).
\end{equation}
Together with the K\"unneth formula in cohomology, this means that the cohomology ring of $BG_{\text{SM}}$ is generated by the Chern classes associated with each factor of the gauge group,
\begin{equation} \label{cohomology ring for BG SM}
    H^{\bullet}\left(BG_{\text{SM}};\Z\right) \cong \Z\left[x,c_2^\prime,c_2,c_3\right],
\end{equation}
where $x\in H^2\left(BG_{\text{SM}};\Z\right)$ indicates the first Chern class associated with the $U(1)$ factor, $c_2^\prime\in H^4\left(BG_{\text{SM}};\Z\right)$ indicates the second Chern class of $SU(2)$, and $c_2\in H^4\left(BG_{\text{SM}};\Z\right)$ and $c_3\in H^6\left(BG_{\text{SM}};\Z\right)$ indicate the second and third Chern classes respectively of the $SU(3)$ factor.
We thus have the following low dimension cohomology groups
\begin{nalign}
    H^0\left(BG_{\text{SM}};\Z\right) &\cong \Z,\\
    H^2\left(BG_{\text{SM}};\Z\right) &\cong \Z,\\
    H^4\left(BG_{\text{SM}};\Z\right) &\cong \Z^3,\\ 
    H^6\left(BG_{\text{SM}};\Z\right) &\cong \Z^4,
\end{nalign}
with all cohomology groups in odd degrees vanishing.
Because of this, and because these groups are all torsion-free, there is a (non-canonical) isomorphism
\begin{equation}
H_{2k}\left(BG_{\text{SM}};\Z\right) \cong H^{2k}\left(BG_{\text{SM}};\Z\right),
\end{equation}
yielding the homology groups that we need to populate the entries of the second page of the AHSS relevant for computing the bordism groups $\Omega^{\text{Spin}}_{d}(BG_{\text{SM}})$ up to $d=5$, since we know that
\begin{equation}
E^2_{p,q}=H_p(BG_{\text{SM}};\Omega^{\text{Spin}}_q(\text{pt}))=H_p(BG_{\text{SM}};\Z) \otimes \Omega^{\text{Spin}}_q(\text{pt}),
\end{equation}
where the bordism groups of a point $\Omega^{\text{Spin}}_q(\text{pt})$ are as listed in Eq. (\ref{eq:bordism-table}). The entries of the second page are shown in Fig.~\ref{fig:AHSS SM}.

Since the action of the Steenrod square on the generators of $H^\bullet(BSU(n);\Z/2)$, which are the universal Chern classes, is given by the formula~\cite{Garcia-Etxebarria:2018ajm}
\begin{equation}
  \text{Sq}^2\left( c_i\right) = (i-1)\,c_{i+1}\nn
\end{equation}
the Steenrod square action on each of the generators of the cohomology ring (\ref{cohomology ring for BG SM}) is then given by
\begin{nalign} \label{SM Sq}
    \text{Sq}^2(x) &= x^2,\\
    \text{Sq}^2(c_2^\prime)&=0,\\
    \text{Sq}^2(c_2)&=c_3,\\
    \text{Sq}^2(c_3)&=0,
\end{nalign}
where $x^2$ is a shorthand notation for $x \cup x$, the cup product of cohomology classes. This follows from the third line of Eq. \eqref{eq: Cartan} and naturality of the Steenrod squares, as discussed at the end of \S \ref{sec:methodology}. We see from Fig.~\ref{fig:AHSS SM} that there is only a single entry on the diagonal $p+q=5$ which is thus relevant to the computation of $\Omega_5^{\text{Spin}}(BG_{\text{SM}})$, and that is $E^2_{4,1}$. We need to compute what this stabilises to, so we begin by turning to the third page, which requires us to compute the differentials labelled $\beta$ and $\gamma$ in Fig.~\ref{fig:AHSS SM}.

Using the Steenrod squares (\ref{SM Sq}), together with Eqs. (\ref{steenrod 1 to 2}) and the fact that $\Omega^{\text{Spin}}_1(\text{pt})=\Omega^{\text{Spin}}_2(\text{pt})=\Z/2$, we have that the differential labelled $\beta$ in Fig.~\ref{fig:AHSS SM} is the dual of the Steenrod square
\begin{nalign}
    \text{Sq}^2: H^2\left(BG_{\text{SM}};\Z/2\right)&\longrightarrow H^4\left(BG_{\text{SM}};\Z/2\right)\\
    x &\mapsto x^2.
\end{nalign}
Let us denote the generators of $E^2_{4,1}\cong (\Z/2)^3$ as $\widetilde{x^2}$, $\widetilde{c_2^\prime}$, and $\widetilde{c_2}$, which are dual to the generators $x^2$, $c_2^\prime$, $c_2\in H^4\left(BG_{\text{SM}};\Z/2\right)$ by the Kronecker pairing (denoted $\left< \cdot, \cdot \right>$) between homology and cohomology.
Then we see that
\begin{nalign}
\left<\widetilde{\text{Sq}^2}\widetilde{x^2},x\right> &= \left<\widetilde{x^2},x^2\right> = 1,\\
\left<\widetilde{\text{Sq}^2}\widetilde{c_2^\prime},x\right> &= \left<\widetilde{c_2^\prime},x^2\right> = 0,\\
\left<\widetilde{\text{Sq}^2}\widetilde{c_2},x\right> &= \left<\widetilde{c_2},x^2\right> = 0,\\
\end{nalign}
where $\widetilde{\text{Sq}^2}$ denotes the dual Steenrod square. 
Hence, the kernel of $\beta$ is $\ker\beta \cong (\Z/2)^2$, generated by $\widetilde{c_2^\prime}$ and $\widetilde{c_2}$.

The differential labelled $\gamma$ in Fig.~\ref{fig:AHSS SM} is the composition of the dual Steenrod square and the reduction mod 2:
\begin{equation}
    \gamma : \Z^4 \xrightarrow{\mod 2} (\Z/2)^4 \xrightarrow{\widetilde{\text{Sq}^2}} (\Z/2)^3,
\end{equation}
where the relevant Steenrod square is 
\begin{nalign} \label{gamma steenrod square SM}
\text{Sq}^2:H^4\left(BG_{\text{SM}};\Z/2\right)&\longrightarrow H^6\left(BG_{\text{SM}};\Z/2\right)\\
x^2 & \mapsto 2x^3 = 0 \mod 2,\\
c_2^\prime & \mapsto 0,\\
c_2 & \mapsto c_3,
\end{nalign}
where to deduce $x^2 \mapsto 2x^3$ we have used Cartan's formula \eqref{eq: Cartan} and the fact that $\text{Sq}^1(x) = 0$ as $H^3$ is trivial. Again using the Kronecker pairing, we deduce that $\widetilde{\text{Sq}^2}$ kills $\widetilde{x^3}$, $\widetilde{c_2\cup x}$, $\widetilde{c_2^\prime\cup x}$, and sends $\widetilde{c_3}$ to $\widetilde{c_2}$. Therefore $\Ima \gamma \cong \Z/2$, generated only by $\widetilde{c_2}$.
We can then take the homology with respect to the differentials $\beta$ and $\gamma$ to turn the page of the AHSS and deduce the $(4,1)$ element of the third page,
\begin{equation} \label{E3 41 SM}
    E^3_{4,1} = \frac{\ker \beta}{\Ima \gamma} \cong \frac{(\Z/2)^2}{\Z/2} \cong \Z/2.
\end{equation}
Since the entries in every odd column vanish, there are no non-trivial differentials on the third page, and so we can turn to the fourth page with $E^4_{p,q}=E^3_{p,q}$ for all $(p,q)$.

On the fourth page the only differential relevant to computing $\Omega_5^{\text{Spin}}(BG_{\text{SM}})$ is $d^4:E^4_{4,1}\rightarrow E^4_{0,5}$, which is
a homomorphism from $\Z/2$ to $\Z$ and is thus trivial. So the $(4,1)$ entry stabilises to $E_\infty^{4,1}\cong \Z/2$, and since this is the only non-zero element on the $p+q=5$ diagonal it follows that 
\begin{equation}
    \Omega^{\Spin}_5\left(BG_{\text{SM}}\right) \cong \Z/2,
\end{equation}
where we can identify the potential global anomaly in this theory with the Witten anomaly associated to the $SU(2)$ factor.

To see that this must be the case, consider a theory with gauge group $G_{\text{SM}}$ and a single fermion transforming as a doublet under $SU(2)$ and a singlet under both $SU(3)$ and hypercharge. Using the Dai--Freed prescription for the fermionic partition function one obtains an anomalous theory because $\exp 2\pi i \eta = -1$ on $S^4\times S^1$. This must therefore correspond to the non-trivial class in $\Omega^{\Spin}_5\left(BG_{\text{SM}}\right)$.

%\medskip
%%%%%%%%%%%%%%%%%%%%%%%

We can continue to compute the bordism groups of $BG_{\text{SM}}$ in lower degrees in a similar fashion. From Fig.~\ref{fig:AHSS SM} we can immediately read off
\begin{equation}
\Omega^{\Spin}_0\left(BG_{\text{SM}}\right) \cong \Z, \qquad \text{and} \qquad \Omega^{\Spin}_1\left(BG_{\text{SM}}\right) \cong \Z/2,
\end{equation}
and it is straightforward to show that 
\begin{equation}
    \Omega^{\Spin}_2\left(BG_{\text{SM}}\right) \cong \Z \times \Z/2,
\end{equation}
Next, to compute $\Omega^{\Spin}_3\left(BG_{\text{SM}}\right)$, we need the differential
\begin{equation}
    \alpha : \Z^3 \xrightarrow{\mod 2} (\Z/2)^3 \xrightarrow{\widetilde{\text{Sq}^2}} \Z/2,
\end{equation}
as well as the map $d^2_{2,1}:\Z/2 \rightarrow \Z/2$.
The dual Steenrod square is precisely the same as for the map $\beta$, which maps $\widetilde{x^2}\mapsto \widetilde{x}$, and the other generators to zero, so we have that $\Ima \alpha=\Z/2$. Then, we do not need to compute the map $d^2_{2,1}$ to deduce that its kernel must be $\Z/2$, because we know that $\Ima \alpha \subset \ker d^2_{2,1}$. Hence, taking the homology, we deduce that $E^\infty_{2,1}=0$. All elements on the $p+q=3$ diagonal thus stabilise to zero and we have that 
\begin{equation}
    \Omega^{\Spin}_3\left(BG_{\text{SM}}\right) =  0.
\end{equation}
To compute $\Omega^{\Spin}_4\left(BG_{\text{SM}}\right)$, we know from above that the map $\beta$ into $E^2_{2,2}$ has image $\Ima \beta \cong \Z/2$, generated by the element $\widetilde{x}\in H_2(BG_{\text{SM}};\Z/2)$. The map out of $E^2_{2,2}$ is to zero and so its kernel is $\Z/2$; turning to the next page, this element therefore stabilises at $\frac{\Z/2}{\Z/2} = 0$. More care is required to deduce $\ker \alpha$, as follows. We have that $\widetilde{c_2^\prime}$ and $\widetilde{c_2}$ certainly map to zero, where note that the elements $\widetilde{x^2}$, $\widetilde{c_2^\prime}$, and $\widetilde{c_2}$ are here valued in integral homology (rather than in homology with coefficients in $\Z/2$).
Thus, while $\widetilde{x^2}\in H^4\left(BG_{\text{SM}};\Z\right)$ maps to the non-zero element $\widetilde{x}\in H^2\left(BG_{\text{SM}};\Z/2\right)$, the element $2\widetilde{x^2}\in H^4\left(BG_{\text{SM}};\Z\right)$ maps to zero in $H^2\left(BG_{\text{SM}};\Z/2\right)$. Hence, the map $\alpha$ has a kernel $\ker \alpha \cong \Z^3$ (which may look strange given its image is non-zero), and so we deduce $E^\infty_{4,0}\cong \Z^3$. Given also that $E^\infty_{0,4}\cong \Z$, we compute
\begin{equation}
    \Omega^{\Spin}_4\left(BG_{\text{SM}}\right) \cong \Z^4,
\end{equation}
thus concluding our computation of the bordism groups $\Omega^{\text{Spin}}_{d\leq 5}(BG_{\text{SM}})$ for the SM gauge group without a quotient. This result, along with others, is summarized in Table~\ref{summary bordism}.

\begin{table}[h]
\begin{center}
  \begin{tabular}{c|cccccc}

    & \multicolumn{6}{c}{$\Omega^{\Spin}_d(BG)$} \\
  $G$ & 0 & 1 & 2 & 3 & 4 & 5\\
  \hline
  $U(1) \times SU(2)\times SU(3)$ & $\Z$ & $\Z/2$ & $\Z\times\Z/2$ & $0$ & $\Z^4$ & $\Z/2$\\
  $(U(1) \times SU(2)\times SU(3))/\Gamma_2$ & $\Z$ & $\Z/2$ & $\Z\times\Z/2$ & $0$ & $\Z^4$ & $0$\\
  $(U(1) \times SU(2)\times SU(3))/\Gamma_3$ & $\Z$ & $\Z/2$ & $\Z\times\Z/2$ & $0$ & $\Z^4$ & $\Z/2$\\
  $(U(1) \times SU(2)\times SU(3))/\Gamma_6$ & $\Z$ & $\Z/2$ & $e(\Z/3,\Z\times\Z/2)$ & $0$ & $e(\Z/3,e(\Z/3,\Z^4))$ & $0$\\
\end{tabular}
\vspace{5mm}
\caption{Summary of results from our bordism computations for the four possible SM gauge groups. We tabulate the bordism groups in degrees zero through five.} \label{summary bordism}
\end{center}
\end{table}

\subsection{\texorpdfstring{$\Omega_5^{\text{Spin}}(B(G_{\text{SM}}/\Gamma_2))$}{Bordism group of Z2 quotient} }
\label{sec:Z2 SM quotient}

We now turn to compute the bordism groups for the variants of the SM involving quotients of $G_{\text{SM}}$ by discrete subgroups of its center, as listed in Eq. (\ref{SM gauge group}).
Recall from \S \ref{sec:hypercharge} that
\begin{equation}
\frac{G_{\text{SM}}}{\Gamma_2} \cong SU(3)\times U(2).
\end{equation}
Hence $B(G_{\text{SM}}/\Gamma_2)=BU(2)\times BSU(3)$ using (\ref{eq:BG}). This is useful, because the cohomology ring of the classifying space of the groups $U(n)$ is well-known. 

\begin{figure}[h]
\includegraphics[width=0.55\textwidth]{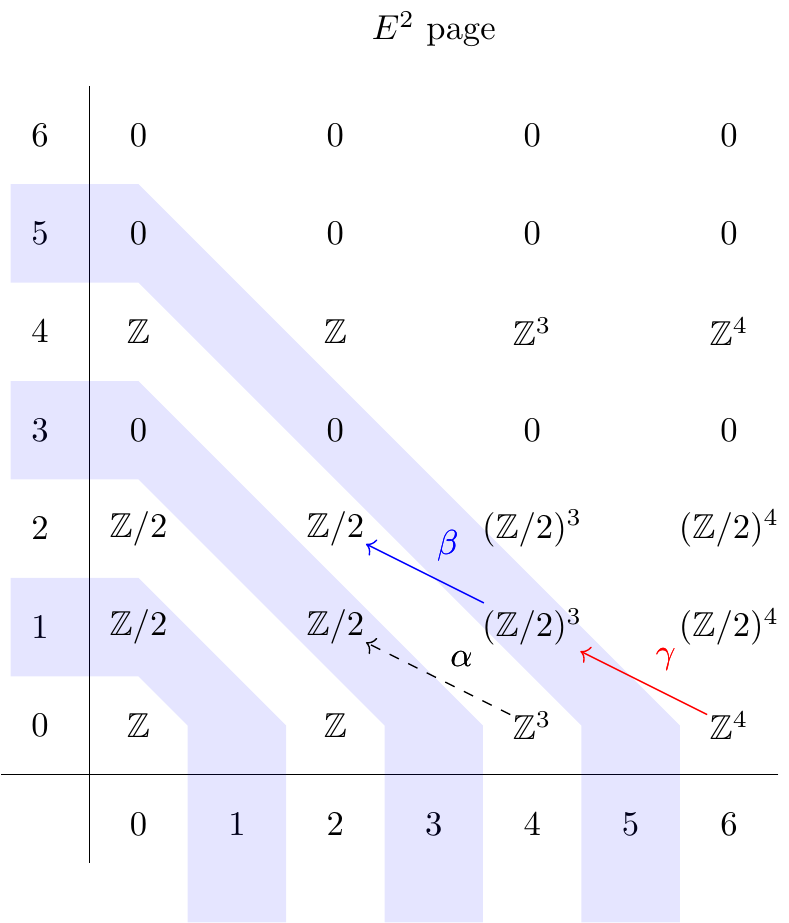}
\centering
  \caption{The $E^2$ page of the Atiyah-Hirzebruch spectral sequence for $G=U(2)\times SU(3)$, with differentials relevant to the computation of the fourth and fifth bordism groups labelled.}
  \label{fig:AHSS U(2)xSU(3) E2}
\end{figure}

Using the usual fibration
$\text{pt} \longrightarrow B(G_{\text{SM}}/\Gamma_2)\longrightarrow B(G_{\text{SM}}/\Gamma_2)$,
the second page of the AHSS is given by
$E^2_{p,q} = H_p\left(BU(2)\times BSU(3);\Omega^{\text{Spin}}_q(\text{pt})\right)$, as shown in figure \ref{fig:AHSS U(2)xSU(3) E2}.
Recall that the relevant cohomology rings are
\begin{nalign}
  H^\bullet\left(BSU(3)\right) =& \Z[c_2,c_3]\\
  H^\bullet\left(BU(2)\right) =& \Z[c^\prime_1,c^\prime_2]
\end{nalign}
where $c_i, c^\prime_i$ are the $i$th Chern classes (which are cohomology classes in degree $2i$) for $SU(3)$ and $U(2)$, respectively. Thus, we have the integral cohomology groups
\begin{nalign}
    H^0\left(B(G_{\text{SM}}/\Gamma_2);\Z\right) &\cong \Z, \\
    H^2\left(B(G_{\text{SM}}/\Gamma_2);\Z\right) &\cong \Z, \quad \text{generated by} \quad c^\prime_1, \\
    H^4\left(B(G_{\text{SM}}/\Gamma_2);\Z\right) &\cong \Z^3, \quad \text{generated by} \quad c^{\prime 2}_1,\ c^\prime_2,\ c_2,\\ 
    H^6\left(B(G_{\text{SM}}/\Gamma_2);\Z\right) &\cong \Z^4, \quad \text{generated by} \quad c^{\prime 3}_1,\ c^\prime_1 c^\prime_2,\ c^\prime_1 c_2,\ c_3.
\end{nalign}
Again, because these are torsion-free and the cohomology groups all vanish in odd degrees, we deduce from these the integral homology groups, 
\begin{equation}
H_{2k}\left(B(G_{\text{SM}}/\Gamma_2);\Z\right) \cong H^{2k}\left(B(G_{\text{SM}}/\Gamma_2);\Z\right).
\end{equation}

Thus far, this appears superficially identical to the case of no discrete quotient considered above, and indeed the second page of the AHSS is populated by the same groups; however, the action of the Steenrod squares is subtly different, meaning the action of the differentials (and, specifically, the maps $\alpha$, $\beta$, and $\gamma$) is not necessarily the same as above. It turns out that an important difference shall be in the map $\gamma$. In particular, since the action of the Steenrod square on the generators $c_i$ of $H^\bullet(BU(n);\Z/2)\cong \Z/2[c_1,\ldots,c_n]$ is  given by~\cite{borel1953groupes}
\begin{equation}
  \label{eq: Sq on BU}
  \text{Sq}^2(c_i) = c_1\cup c_i + (i-1)c_{i+1},
\end{equation}
we have that its action on the generators of the cohomology ring of $B(U(2)\times SU(3))$ is
\begin{nalign} \label{SM Sq 2}
    \text{Sq}^2(c_1^\prime) &= c_1^{\prime 2},\\
    \text{Sq}^2(c_2^\prime)&=c_1^\prime \cup c_2^\prime,\\
    \text{Sq}^2(c_2)&=c_3,\\
    \text{Sq}^2(c_3)&=0.
\end{nalign}
Notice the second line in particular, to be contrasted with the second line in Eq. (\ref{SM Sq}). As before, this follows from naturality of the Steenrod square.

The differentials relevant to the calculation of $\Omega^{\Spin}_4\left(B(G_{\text{SM}}/\Gamma_2)\right)$ and $\Omega^{\Spin}_5\left(B(G_{\text{SM}}/\Gamma_2)\right)$ are again given by
\begin{nalign} \label{differentials SM quotients}
  \alpha &= \widetilde{\text{Sq}^2} \circ \rho,\\
  \beta &= \widetilde{\text{Sq}^2},\\
  \gamma &= \widetilde{\text{Sq}^2}\circ \rho,
\end{nalign}
where $\rho$ denotes reduction modulo 2.
Since $\text{Sq}^2:H^2\rightarrow H^4$ maps $c^\prime_1\mapsto c^{\prime 2}_1$, we see that both $\alpha, \beta$ map $\widetilde{c^{\prime 2}_1}\mapsto \widetilde{c^\prime_1}$ and others to zero. Moreover, $\alpha$ maps $2\widetilde{c^{\prime 2}_1}$ to zero. So we have, using similar arguments as before, that
\begin{equation}
\ker \alpha \cong \Z^3,\quad \Ima \alpha \cong \Z/2,\qquad \ker \beta = (\Z/2)^2, \quad \Ima \beta \cong \Z/2,
\end{equation}
which is as it was in the previous case. 

We now turn to the map $\gamma$. The relevant Steenrod square is here
\begin{nalign}
\text{Sq}^2:H^4\left(B(G_{\text{SM}}/\Gamma_2);\Z/2\right)&\longrightarrow H^6\left(B(G_{\text{SM}}/\Gamma_2);\Z/2\right)\\
c^{\prime 2}_1 & \mapsto 2c^{\prime 3}_1 \equiv 0 \mod 2,\\
c^\prime_2 & \mapsto c^\prime_1 \cup c^\prime_2,\\
c_2 & \mapsto c_3,
\end{nalign}
where the third line should be contrasted with that in Eq. (\ref{gamma steenrod square SM}).
So $\gamma$ maps $\widetilde{c^\prime_1 \cup c^\prime_2}\mapsto \widetilde{c^\prime_2}$ and $\widetilde{c_3}\mapsto \widetilde{c_2}$, while mapping other generators to zero. This gives $\Ima \gamma \cong (\Z/2)^2$. Then
\begin{equation}
E^3_{4,1} = \frac{\ker \beta}{\Ima \gamma} = 0,
\end{equation}
to be contrasted with the non-zero result in Eq. (\ref{E3 41 SM}). Thus, this entry stabilises, and there are no non-zero entries on the diagonal $p+q=5$ of the last page of this AHSS. Hence, we deduce
\begin{equation}
\Omega^{\text{Spin}}_5(B(G_{\text{SM}}/\Gamma_2))=0,
\end{equation}
and thus that this version of the SM has no global anomalies, no matter what the fermion content.
One can compute the bordism groups in lower degrees using the same methods as in the previous example, and one finds no other differences in the results, which are again recorded in Table~\ref{summary bordism}.

We thus arrive at a seemingly curious result; there are no global anomalies in this version of the SM, for arbitrary fermion content. The reader might wonder what has happened to the Witten anomaly, and the condition that there must be an even number of $SU(2)$ doublets in the theory. We discuss the resolution to this puzzle (which also occurs in the case $G=G_{\text{SM}}/\Gamma_6$) in \S \ref{sec:interplay}. For now, it might be useful to remark on what goes wrong with the argument of the previous Section, in which we considered a theory with a single fermion in the spin-$\frac{1}{2}$ representation of $SU(2)$ (and a singlet under both $SU(3)$ and $U(1)$), and claimed $\exp 2\pi i \eta = -1\neq 1$ on $S^1\times S^4$. We cannot use such an argument when $G=G_{\text{SM}}/\Gamma_2$, because the hypercharge constraints presented in \S \ref{sec:hypercharge} mean there is no such representation of the gauge group, because any $SU(2)$ doublet fermion must have {\em odd} (and thus non-zero) hypercharge. We must then take care to ensure that {\em local} anomalies associated with hypercharge cancel, before we turn to the global anomalies. We return to this issue in \S \ref{sec:interplay}.

%%%%%%%%%%%%%%%%%

\subsection{\texorpdfstring{$\Omega_5^{\text{Spin}}(B(G_{\text{SM}}/\Gamma_3))$}{Bordism group of Z3 quotient}}
\label{sec:SM Z3}
Our approach for tackling this variant of the SM is qualitatively very similar to that employed for the $\Z/2$ quotient in the previous Subsection. Recall from \S \ref{sec:hypercharge} that the gauge group here may written as
\begin{equation} \label{eq: Z3 quotient xxx}
\frac{G_{\text{SM}}}{\Gamma_3}=\cong U(3)\times SU(2).
\end{equation}
One may tackle this variant of the SM using the same methods employed for the $\Z/2$ quotient in the previous Subsection. Thus, to avoid repetition, we relegate the calculations for this gauge group to Appendix \ref{App: alt Z3}. The upshot is that we find
\begin{equation}
\Omega_5^{\text{Spin}}(B(G_{\text{SM}}/\Gamma_3))=\Z/2,
\end{equation}
corresponding to the Witten anomaly associated with the $SU(2)$ factor in (\ref{eq: Z3 quotient xxx}). The lower-degree bordism groups are tabulated in Table~\ref{summary bordism}.

For this gauge group, an alternative fibration exists which we can also use to compute the bordism groups, based on the Puppe sequence. Reassuringly, using this other fibration yields the same bordism groups, and we include the details of both methods in Appendix \ref{App: alt Z3}. We will need to employ such a Puppe-induced fibration shortly in \S \ref{sec: SM Z6} to compute the bordism groups of $B(G_{\text{SM}}/\Gamma_6)$.

\subsection{\texorpdfstring{$\Omega_5^{\text{Spin}}(B(G_{\text{SM}}/\Gamma_6))$}{Bordism group of Z6 quotient} }
\label{sec: SM Z6}

The $\Z/6$ quotient in the case $G=G_{\text{SM}}/\Gamma_6$ is generated by the element $\xi$ given by \eqref{eq:xi},
and there is no straightforward way to write the group $G_{\text{SM}}/\Gamma_6$ as a product, as we did in the previous two cases. This means a direct attempt to use the AHSS to compute the bordism groups of $G_{\text{SM}}/\Gamma_6$ seems unlikely to work, given we do not know how the differentials on the second page act.

Instead, we consider the following fibration\footnote{We note, to avoid confusion, that there also exists a fibration of the group $U(2)\times SU(3)$ over $U(2)\times PSU(3)$ (which cannot be the gauge group of the Standard Model because $PSU(3)$ does not admit a triplet representation) with the same homotopy fibre. While this fibration would be written using the same notation as \eqref{eq: fibration for finite quotient}, the maps are, of course, different.}
\begin{equation} \label{eq: fibration for finite quotient}
\Z/3 \longrightarrow U(2)\times SU(3) \longrightarrow G_{\text{SM}}/\Gamma_6.
\end{equation}
This induces the fibration $B(\Z/3) \rightarrow B(U(2)\times SU(3)) \rightarrow B(G_{\text{SM}}/\Gamma_6)$, which turns into the following, more useful, fibration after we invoke the Puppe sequence (we here follow a similar strategy to that used in Ref.~\cite{gu2016cohomology}):
\begin{equation} \label{eq:practical fibration for Z_6 quotient}
B\left(U(2)\times SU(3)\right) \longrightarrow B(G_{\text{SM}}/\Gamma_6) \longrightarrow K\left(\Z/3,2\right),
\end{equation}
where $K\left(\Z/3,2\right)=B(B(\Z/3))$ is an Eilenberg-Maclane space.

\begin{figure}[h]
\includegraphics[width=0.5\textwidth]{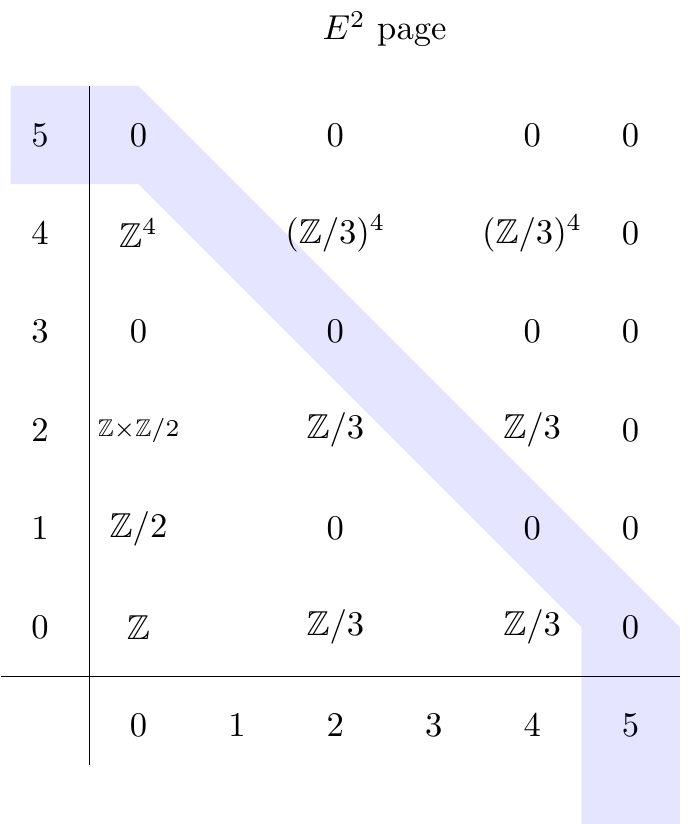}
\centering
  \caption{The second page of the Atiyah-Hirzebruch spectral sequence corresponding to the fibration (\ref{eq:practical fibration for Z_6 quotient}). The entries relevant to the computation of $\Omega_5^{\text{Spin}}(BG_{\text{SM}}/\Gamma_6)$ are highlighted, all of which vanish already on the second page.
  }
  \label{fig: AHSS E2 SM/Z6}
\end{figure}

The second page of the AHSS associated with this fibration is given by
\begin{equation} \label{AHSS pg 2 Z6 quotient}
E^2_{p,q} = H_p\left(K\left(\Z/3,2\right);\Omega^{\text{Spin}}_q(B(U(2)\times SU(3)))\right).
\end{equation}
While this may look like a rather unwieldy expression, note that the bordism groups $\Omega^{\text{Spin}}_q(B(U(2)\times SU(3)))$ are precisely those that we have already computed in our study of global anomalies for the case $G=G_{\text{SM}}/\Gamma_2$, as recorded in the second line of Table~\ref{summary bordism}. These groups only feature factors of $\Z$ and $\Z/2$, and the homology groups of the Eilenberg-Maclane space $K(\Z/3,2)$ valued in $\Z$ and $\Z/2$ are~\cite{breen2016derived}
\begin{equation} \label{eq: homology of KZ3}
\def\arraystretch{1.5}
\arraycolsep=4pt
\begin{array}{c|cccccc}
i & 0 & 1 & 2 & 3 & 4 & 5   \\
\hline
H_i(K(\Z/3,2),\Z) & \Z & 0 & \Z/3 & 0 & \Z/3 & 0   \\
H_i(K(\Z/3,2),\Z/2) & \Z/2 & 0 & 0 & 0  & 0 & 0. 
\end{array}
\end{equation}
We can thence compute all the entries (\ref{AHSS pg 2 Z6 quotient}) in the second page of the AHSS. These are shown in 
Fig.~\ref{fig: AHSS E2 SM/Z6}.

Somewhat fortunately (for the sake of being able to perform the computation), all the entries on the $p+q=5$ diagonal relevant for the computation of $\Omega_5^{\text{Spin}}(BG_{\text{SM}}/\Gamma_6)$ vanish already on the second page. This is just as well, because for this fibration we do not know any formulae for the action of the differentials (with which to turn to the next page) in terms of Steenrod squares (or indeed any other operation on (co)homology).\footnote{Note that the similar-looking fibration $\Z/2 \longrightarrow U(3)\times SU(2) \longrightarrow G_{\text{SM}}/\Gamma_6$ does not yield such simplifications, and so cannot be used to compute the relevant bordism group because there are unknown differentials on the second page.  This is roughly because the homology of $K(\Z/2,2)$ is `more complicated' than that of $K(\Z/3,2)$. } We thus conclude that
\begin{equation}
\Omega^{\text{Spin}}_5(B(G_{\text{SM}}/\Gamma_6))=0.
\end{equation}
Since all relevant homomorphisms are trivial, all entries $E_{p,q}$ with $p+q<5$ stabilise on the second page. We can then compute the remaining bordism groups with degree lower than  $5$ without ambiguities apart from $\Omega^{\Spin}_2(B(G_{\text{SM}}/\Gamma_6))$ and $\Omega^{\Spin}_4(B(G_{\text{SM}}/\Gamma_6))$ due to non-splitting extensions. They are given by
\begin{align}
  \Omega^{\Spin}_2(B(G_{\text{SM}}/\Gamma_6)) &= e(\Z/3,\Z\times\Z/2),\nn\\
  \Omega^{\Spin}_4(B(G_{\text{SM}}/\Gamma_6)) &= e(\Z/3,e(\Z/3,\Z^4)). \label{eq:AHSS ambiguities}
\end{align}
The notation $e(A,B)$ denotes a group extension of $A$ by $B$, that is, a group that fits into the following short exact sequence
\begin{equation}
  0\longrightarrow B \longrightarrow e(A,B) \longrightarrow A \longrightarrow 0.
  \label{eq: group extension defn}
\end{equation}
We tabulate our results in Table~\ref{summary bordism}.

\medskip

{\em Note added}: since this article appeared in preprint form, the Adams spectral sequence has been used to resolve the ambiguities we found (using the AHSS) in Eq. (\ref{eq:AHSS ambiguities})~\cite{Wan:2019fxh}. It was therein found that 
\begin{equation}
  \Omega^{\Spin}_2(B(G_{\text{SM}}/\Gamma_6)) = \Z\times\Z/2. \label{eq:WanWang1}
\end{equation}
Comparing with our result (\ref{eq:AHSS ambiguities}), this corresponds to the non-trivial extension
\begin{equation}
0 \longrightarrow \Z \times \Z/2 \longrightarrow \Z \times \Z/2 \longrightarrow \Z/3 \longrightarrow 0,
\end{equation}
where the first map is multiplication by 3 on the first factor and the identity on the second. In Ref.~\cite{Wan:2019fxh} it was also found that
\begin{equation}
  \Omega^{\Spin}_4(B(G_{\text{SM}}/\Gamma_6)) = \Z^4, \label{eq:WanWang2}
\end{equation}
also corresponding to a non-trivial solution to the extension problem (\ref{eq:AHSS ambiguities}).

\subsection{Interplay between global and local anomalies} \label{sec:interplay}

It is interesting that there are no possible global anomalies in the cases with quotients by $\Z/2$ and $\Z/6$, whereas in the case of a quotient by $\Z/3$ (or the case with no quotient at all) there is a $\Z/2$ global anomaly which we have identified with the familiar Witten anomaly associated with the $SU(2)$ factor. 

This might at first appear puzzling. We know that cancellation of the Witten anomaly in an $SU(2)$ gauge theory, and in the SM, requires $n_L-n_R=0$ mod 2 if there are $n_L$ ($n_R$) left-handed (right-handed) fermions in $SU(2)$ doublets. More generally, the Witten anomaly receives contributions from any fermions in $SU(2)$ representations with isospin $2r+1/2$, $r\in \Z$. Does the fact that we have computed that there are no such conditions for global anomaly cancellation in two variants of the SM mean that in these cases we can dispense with Witten's condition, and consider extensions of the SM with odd numbers of $SU(2)$ doublets? The answer is no, due to a subtle interplay between global and local anomaly cancellation, which we now describe. 

The key point is that taking discrete quotients of $G_{\text{SM}}$ changes the set of representations that fermions can carry, since every fermion must be in a {\em bona fide} representation of the group $G$. This leads to constraints on the possible hypercharges for fermions transforming as electroweak doublets. As we derived in \S \ref{sec:hypercharge}, when we quotient $G_{\text{SM}}$ by $\Z/2$ or $\Z/6$, any field transforming in the $(j,q)$ representation of the $SU(2)\times U(1)$ factor must satisfy the isospin-charge relation
\begin{equation} \label{eq: U2 charge relations again}
q=2j \text{~mod~} 2.
\end{equation}
Of course, one is free to perform an overall rescaling of all the $U(1)$ charges in the theory, so the precise statement is that there must exist a normalisation of the $U(1)$ gauge coupling such that the charge constraints (\ref{eq: U2 charge relations again}) are possible. We assume such a normalisation for the $U(1)$ charges in the following.\footnote{Note that the local anomaly cancellation equations are homogeneous polynomials in rational charges, and thus are properly defined on a projective rational variety; thus, we are free to fix an overall normalisation as we wish.
}

Now consider the cancellation of local anomalies. Suppose we have $N_j$ fermions transforming in the $SU(2)$ representation with isospin $j$, and that these have charges denoted $\{q_j^{(a)}\}$, where $a=1,\dots N_j$, and $q_j^{(a)}=2j \text{~mod~} 2$. We assume that all fermions have left-handed chirality. The $SU(2)^2\times U(1)$ anomaly coefficient is then proportional to
\begin{equation} \label{eq:SU22 U1}
\sum_j T(j)  \sum_{a=1}^{N_j} q_j^{(a)}=0,
\end{equation}
where the sum over $j$ is over the different values of isospin, and $T(j)$  denotes the Dynkin index (defined such that $\Tr \left(t^a_j t^b_j  \right)= \frac{1}{2} T(j) \delta_{ab}$, where $\{t^a_j\}$ denotes a basis for $\mathfrak{su}(2)$ in the isospin$-j$ representation), which is given by the formula
\begin{equation}
T(j)=\frac{2}{3}j(j+1)(2j+1).
\end{equation}
This formula implies that $T(j)$ is odd when $j=2r+1/2$, $r\in \Z$, and is even otherwise.

When the anomaly condition (\ref{eq:SU22 U1}) is reduced mod 2, only the contributions to (\ref{eq:SU22 U1}) from isospins $2r+1/2$ remain, since it is only these irreps for which both $T(j)$ and the charges $q_j^{(a)}$ are necessarily odd.
We thus obtain
\begin{equation}
\sum_{j\in 2\Z+1/2} N_j = 0 \mod 2.
\end{equation}
In other words, in the theories with gauge groups $G_{\text{SM}}/\Gamma_2$ or $G_{\text{SM}}/\Gamma_6$, the total number of fermions transforming in isospin $2r+1/2$ representations must be even, in order for the local $SU(2)^2\times U(1)$ anomaly to cancel -- even though there is no global anomaly in either of these cases. This is equivalent to the condition, in the $SU(2)\times U(1)$ case, that the usual Witten anomaly vanishes. This anomaly interplay has been explored more deeply in Ref.~\cite{Davighi:2020bvi}.

%In order for the $U(1)\times SU(2)^2$ anomaly coefficient to cancel requires the sum of hypercharges of fermions that transform in half-integer spin representations of $SU(2)$ must vanish. Thus, since the hypercharges of any $SU(2)$ doublets must all be odd integers in a theory in which the electroweak gauge group is $U(2)$, there must be an even number of such doublets. 

%Note that the local conditions for anomaly cancellation are identical in both the $ U(1)\times  SU(2)$ and $U(2)$ cases, because the ABJ type anomalies are fixed entirely by the Lie algebra of the gauge group. The difference is that in the former case, both even and odd hypercharges are permitted, and so one cannot in general infer a condition on the number of allowed fermions by considering the anomaly cancellation equations mod 2; rather, the requirement that there be an even number of chiral $ SU(2)$ doublets follows from a potential global anomaly. 

%Thus, whether the electroweak gauge group is $U(1)\times SU(2)$ or $U(2)$, in either case we require an even number of fermion doublets for an anomaly-free theory -- even though there is no global anomaly requiring this to be so in the case of $U(2)$.

\section{A generalisation of the  SM} \label{sec:general}

The Standard Model with gauge group $G_{\text{SM}} = SU(3) \times SU(2) \times U(1)$ is the starting point of a 2-parameter family of anomaly-free chiral gauge theories \cite{davidTong, Lohitsiri:2019wpq}. The gauge group for this family of generalised Standard Model theories is
\begin{equation}
  \label{eq: GSM gauge group}
  G_{\text{GSM}} = SU(N) \times Sp(M) \times U(1), \qquad N>2 \text{~and odd}, \; M\geq 1
\end{equation}
It was shown in Ref.~\cite{Lohitsiri:2019wpq} that theories in this family have the same phase structure as the Standard Model when one varies the relative strength between the strong force and the weak force. It is also not far-fetched to assume that this family of theories exhibits similar features in the infrared. This generalisation subjects the Standard Model to the framework of large-$N$ expansion, which could potentially be used to analyse the dynamics of this family of chiral gauge theories perturbatively in a more controlled fashion. %While this family of theories are known to be free of local anomalies, and also Witten anomalies associated with the $Sp(M)$ factor, in this Section we investigate whether there are any global anomalies detected by our more general bordism-based criterion.

The left-handed doublets of fermions that couple to the weak force in the Standard Model now become $2M$-tuplets in the fundamental representation of $Sp(M)$. Since there are $N+1$ chiral fermions in the fundamental representation of $Sp(M)$, we need $N$ to be odd to cancel the $\Z/2$ global anomaly.  In order to have sufficient number of chiral fermions to cancel the local anomalies, the right-handed fermions must proliferate, and we end up with $M$ copies each of right-handed electrons $E_{\alpha}$, right-handed down quarks $D_{\alpha}$, right-handed up quarks $U_{\alpha}$, and right-handed neutrinos $N_{\alpha}$, with $\alpha =1,\ldots, M$. There are also $M$ copies of the Higgs field, $H_\alpha$. The matter content of this generalised theory and its representations under the gauge group $G_{\text{GSM}}$ is given in full in Table \ref{tab: GSM matter}. The simplest case with $M=1$ and $N=3$ gives the Standard Model. 
\begin{table}[h]
\begin{center}
  \begin{tabular}{C||C|C|C}
    & U(1) & Sp(M) & SU(N)\\
    \hline
    Q & +1 & {\bf 2M} & {\bf N}\\
    L & -N & {\bf 2M} & {\bf 1}\\
    D^c_{\alpha} & (2\alpha-1)N-1 & {\bf 1} & \overline{{\bf N}}\\
    U^c_{\alpha} & -(2\alpha-1)N-1 & {\bf 1} & \overline{{\bf N}}\\
    E^c_{\alpha} & 2\alpha N & {\bf 1} & {\bf 1}\\
    N^c_{\alpha} & -(2\alpha-2)N & {\bf 1} & {\bf 1}\\
    H_\alpha & (2\alpha - 1)N & {\bf 2M} & {\bf 1}
  \end{tabular}
\end{center}
\caption{Matter content in the generalised Standard Model. In this table, the boldface characters denote the dimensions of the respective representations, with $\mathbf{2M}$ denoting the fundamental representation of $Sp(M)$ and $\mathbf{N}$ denoting the fundamental of $SU(N)$.  \label{tab: GSM matter}}
\end{table}

The hypercharges given in Table \ref{tab: GSM matter} are chosen so that the theory is free of local anomalies, and the theory is moreover free of Witten anomalies associated with the $Sp(M)$ factor. It is natural to ask whether this generalisation is really consistent for every $(N,M)$ by considering our more general criterion for global anomalies, detected by $\Omega^{\Spin}_5(BG_{\text{GSM}})$. Fortunately, we do not need to repeat our calculation of the spin bordism group for this new gauge group as it is the same as the calculation in \S \ref{sec:no quotient}. To see this, first recall that the relevant entries on the second page of the AHSS are given by
\begin{equation}
  E^2_{p,q} = H_p(BSU(N)\times BSp(M) \times BU(1);\Omega^{\Spin}_q(\text{pt}))\nn
\end{equation}
with $p+q \leq 6$. The K\"unneth theorem for homology then tells us that these entries depend only on $H_r(BSp(M))$ and $H_r(BSU(N))$ with $r\leq 6$. But note that the homology groups in low dimensions of $BSp(M)$ and $BSU(N)$ are given by,
\begin{align}
  H_p(BSp(M);\Z) &= \left\{\Z,0,0,0,\Z,0,0,\ldots\right\},\nn\\
  H_p(BSU(N);\Z) &= \left\{\Z,0,0,0,\Z,0,\Z,\ldots\right\}.\nn
\end{align}
which are the same as those of $SU(2)$ and $SU(3)$, respectively. Therefore, the relevant entries on the second page of the AHSS are still given by Fig. \ref{fig:AHSS SM}. Moreover, the action of the Steenrod square on the generators of lowest degrees of the cohohomology rings of $BSp(M)$ and $BSU(N)$ are the same as in the Standard Model case, giving rise to the same relevant differentials in Fig. \ref{fig:AHSS SM}. The calculation given in \S \ref{sec:no quotient} then goes through unaltered. We then have that
\begin{equation}
  \Omega^{\Spin}_5(BG_{\text{GSM}}) \cong \Z/2
\end{equation}
implying that there is no additional global anomaly except the usual Witten anomaly associated with the $Sp(M)$ factor of the gauge group (for any choice of $M$).
\begin{table}[h!]
\begin{center}
  \begin{tabular}{c|cccccc}
    & \multicolumn{6}{c}{$\Omega^{\Spin}_d(BG)$}\\
  $G$ & 0 & 1 & 2 & 3 & 4 & 5\\
  \hline
  $SU(N)\times Sp(M)\times U(1)$,\ $N>2$ & $\Z$ & $\Z/2$ & $\Z\times\Z/2$ & $0$ & $\Z^4$ & $\Z/2$ 
\end{tabular}
\vspace{5mm}
\caption{The bordism groups pertaining to a generalisation of the SM gauge group.} \label{summary bordism generalised}
\end{center}
\end{table}

\section{Global anomalies in BSM theories} \label{sec:BSM}

In this Section, we  show how to extend these methods to compute whether there are any potential global anomalies in BSM theories, by considering various popular examples. Firstly, we consider extensions of the SM by an arbitrary product of gauged $U(1)$ symmetries (such as in theories featuring heavy $Z^\prime$ gauge bosons).
We then turn to a number of grand unified theories, namely the Pati-Salam model and two trinification models. %, and a five-dimensional theory based on $SO(18)$.

\subsection{Multiple \texorpdfstring{$Z^\prime$}{Z prime} extensions of the SM} \label{sec: many Zprimes}

We consider a four-dimensional gauge theory with gauge group
\begin{equation} \label{G Zprime}
G_{m}\equiv U(1)^m\times SU(2) \times SU(3), \qquad m \geq 2,
\end{equation}
corresponding to an extension of the (usual) SM gauge group by arbitrary $U(1)$ factors, with {\em a priori} arbitrary fermion content. The corresponding $Z^\prime$ bosons in such a theory have been posited to address many phenomenological questions -- for a review, see {\em e.g.} Ref.~\cite{Langacker:2008yv}.
We will compute whether there are potential global anomalies in such a BSM theory.

\begin{figure}[h]
\includegraphics[width=1.04\textwidth]{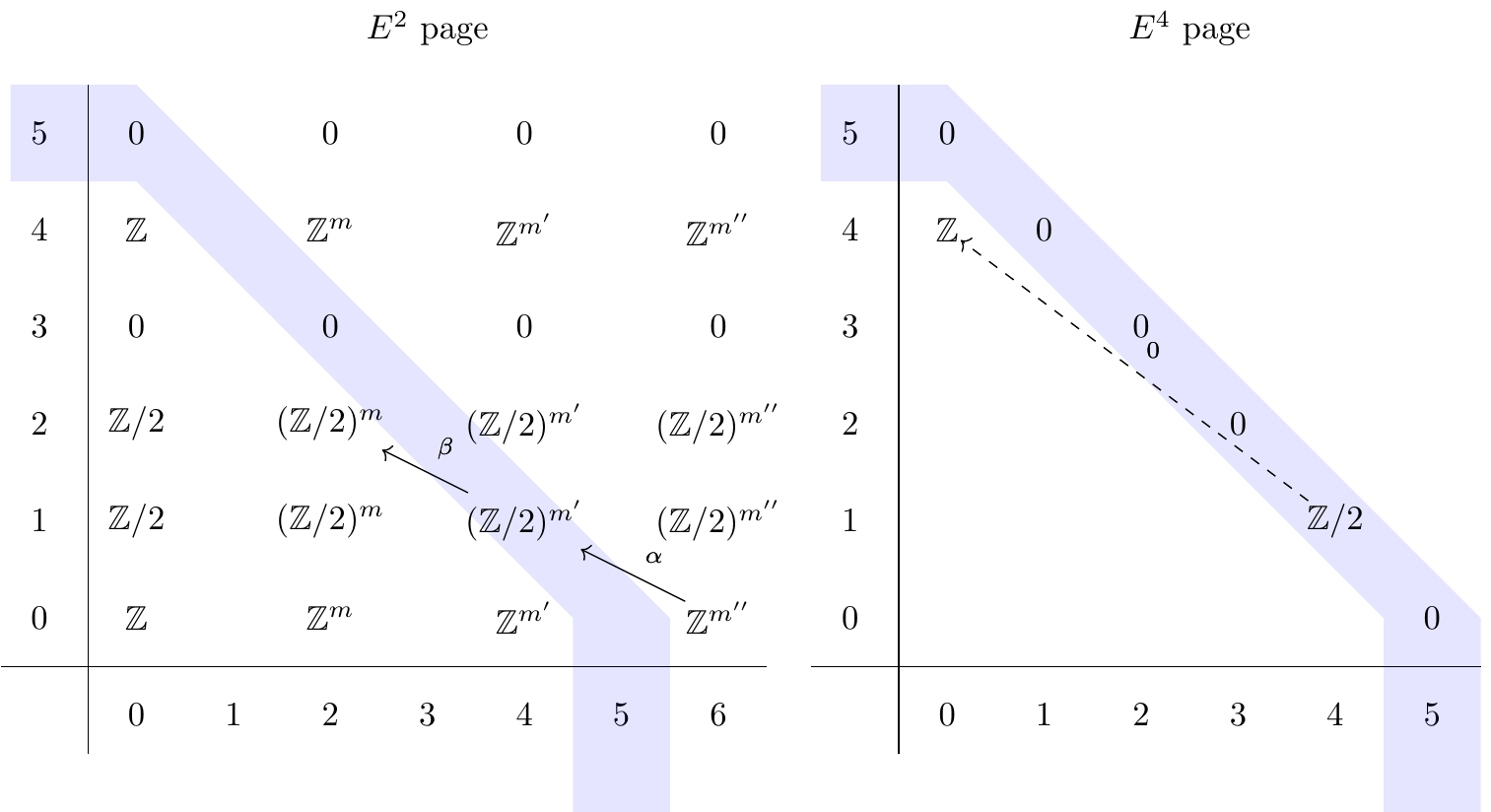}
\centering
  \caption{The $E^2$ and $E^4$ pages  of the Atiyah-Hirzebruch spectral sequence for $G=G_{m}=U(1)^m\times SU(2) \times SU(3)$ with all elements and differentials relevant to the calculation of $\Omega^{\Spin}_5$ highlighted.}
  \label{fig:AHSS U(1)^mxSp(r)xSU(N) E2 5th}
\end{figure}

The cohomology ring for $BG_{m}$ is 
\begin{equation}
    H^{\bullet}\left(BG_{m};\Z\right) \cong \Z\left[x_1,\ldots,x_m,c_2^\prime,c_2,c_3\right],
\end{equation}
where $x_k$ is the first Chern class associated with the $k$th $U(1)$ factor, and the remaining Chern classes are defined as in Eq. (\ref{cohomology ring for BG SM}).
In particular, we have the following low-dimensional cohomology groups
\begin{nalign}
    H^0\left(BG_{m};\Z\right) &\cong \Z,\\
    H^2\left(BG_{m};\Z\right) &\cong \Z^m,\\
    H^4\left(BG_{m};\Z\right) &\cong \Z^{m^\prime},\quad m^\prime = \binom{m+1}{2}+2,\\ 
    H^6\left(BG_{m};\Z\right) &\cong \Z^{m^{\prime\prime}}, \quad m^{\prime\prime} = \binom{m+2}{3}+2m+1,
\end{nalign}
with all cohomology groups in odd degrees vanishing, which of course coincides with the SM case when $m=1$. Again, these groups are isomorphic to the corresponding groups in homology, with which we can deduce the entries $E^2_{p,q}$ of the AHSS, which are shown in Fig.~\ref{fig:AHSS U(1)^mxSp(r)xSU(N) E2 5th}.

We task ourselves here with the computation of $\Omega_5^{\text{Spin}}(BG_{m})$, which measures the potential global anomalies in the four-dimensional gauge theory we are interested in from the point of view of BSM. The relevant entries of the AHSS, lying on the $p+q=5$ diagonal, are highlighted in Fig.~\ref{fig:AHSS U(1)^mxSp(r)xSU(N) E2 5th}. To turn to the third (and thence fourth) page, we thus need to compute the differentials here labelled $\alpha$ and $\beta$. 

This is again similar to the case of the SM considered above. The map $\beta$ is the dual to the Steenrod square
\begin{nalign}
    \text{Sq}^2: H^2\left(BG_{m};\Z/2\right)&\longrightarrow H^4\left(BG_{m};\Z/2\right)\\
    x_i &\mapsto x_i^2,
\end{nalign}
so the kernel of $\beta$ is spanned by $\widetilde{c_2}$, $\widetilde{c_2^\prime}$, and $\widetilde{x_i \cup x_j}$ with $i<j$. Hence  
$ \ker \beta \cong (\Z/2)^{\frac{1}{2}m(m-1)+2}$.
To calculate $\Ima \alpha$, where $\alpha = \widetilde{\text{Sq}^2}\circ\rho$, we first look at the corresponding Steenrod square
\begin{nalign}
\text{Sq}^2: H^4\left(BG_{m};\Z/2\right) &\longrightarrow H^6\left(BG_{m};\Z/2\right)\\
x_i^2 &\mapsto 2x_i^3 \equiv 0 \mod\, 2,\\
x_i x_j &\mapsto x_i^2 x_j + x_i x_j^2,\\
c_2 & \mapsto c_3,\\
c_2^\prime & \mapsto 0.
\end{nalign}
Thus the image of $\widetilde{\text{Sq}^2}$, and also of $\alpha$, is spanned by $\widetilde{c_2}$ and $\widetilde{x_i x_j}$, for $i<j$. Thus $\Ima \alpha \cong (\Z/2)^{\frac{1}{2}m(m-1)+1}$. Taking the quotient then yields
\begin{equation}
E^3_{4,1} = E^4_{4,1} \cong \Z/2.
\end{equation}
On the $E_4$ page (see Fig.~\ref{fig:AHSS U(1)^mxSp(r)xSU(N) E2 5th}) the only relevant differential must be trivial as it is a homomorphism from $\Z/2$ to $\Z$, so the $(4,1)$ entry stabilises to $E^\infty_{4,1}\cong \Z/2$ and it follows that 
\begin{equation}
    \Omega^{\Spin}_5\left(B\left(U(1)^m\times SU(2)\times SU(3)\right)\right) \cong \Z/2,
\end{equation}
where we can again identify the potential global anomaly in this theory with the Witten anomaly associated to the $SU(2)$ factor. Thus we find that there are no potential new global anomalies associated with extending the usual SM gauge group by an arbitrary torus, and indeed by arbitrary fermion content coupled to such a gauge group. There have been a number of recent studies~\cite{Ellis:2017nrp,Allanach:2018vjg,Costa:2019zzy} attempting to classify the space of $U(1)$ extensions of the SM that are free of {\em local} anomalies; here, we show that all such models are automatically free also of global anomalies, provided of course that there is no Witten anomaly associated with $SU(2)$. It is also straightforward to calculate the lower-degree bordism groups for this example, which we simply tabulate in the first line of Table~\ref{summary bordism BSM}. We find that the additional $U(1)$ factors do indeed affect the bordism groups in lower degrees, in particular in degrees two and four.

\begin{table}[h]
\begin{center}
  \begin{tabular}{c|cccccc}
& \multicolumn{6}{c}{$\Omega^{\Spin}_d(BG)$}\\
  $G$ & 0 & 1 & 2 & 3 & 4 & 5\\
  \hline
  \hline
  $U(1)^m\times SU(2)\times SU(3)$ & $\Z$ & $\Z/2$ & $\Z^m\times\Z/2$ & $0$ & $\Z^{3+\frac{1}{2}m(m+1)}$ & $\Z/2$\\
    $SU(4)\times SU(2)_L \times SU(2)_R$ & $\Z$ & $\Z/2$ & $\Z/2$ & $0$ & $\Z^4$ & $(\Z/2)^2$ \\
    $SU(3)_C\times SU(3)_L\times SU(3)_R$ & $\Z$ & $\Z/2$ & $\Z/2$ & 0 & $\Z^4$ & 0\\
    $\dfrac{SU(3)_C\times SU(3)_L\times SU(3)_R}{\Z/3}$ & $\Z$ & $\Z/2$ & $\Z/2 \times \Z/3$ & 0 &  $\Z^4$ or  $\Z^4\times \Z/3$ & 0\\
\end{tabular}
\vspace{5mm}
\caption{Summary of results from our bordism computations of relevance to BSM physics.
  The first row corresponds to theories with multiple $Z^\prime$ bosons,  the second row to a
  Pati-Salam model, and the last two rows to trinification models.} \label{summary bordism BSM}
\end{center}
\end{table}

\subsection{Pati-Salam models}

Here we consider the simplest incarnation (for our purposes) of the Pati-Salam model, in which the SM gauge group is embedded in the larger group
\begin{equation}
\text{PS} \equiv SU(2)_L\times SU(2)_R \times SU(4).
\end{equation}
The cohomology ring for $B(\text{PS})$ is 
\begin{equation}
    H^{\bullet}\left(B(\text{PS});\Z\right) \cong \Z\left[c^L_2,c^R_2,c_2^\prime,c_3^\prime, c_4^\prime\right],
\end{equation}
where $c_2^{L/R}$ denote the second Chern classes of the $SU(2)_{L/R}$ factors, and $c_i^\prime$ denotes the $i$th Chern class of $SU(4)$. A notable difference between this example and all those considered previously is that the second homology group is here vanishing.
This only serves to simplify the computation of the AHSS, and so we choose to omit the details for brevity.
The upshot is that we find
\begin{equation}
\Omega^{\text{Spin}}_{5}\left(B(\text{PS})\right) \cong \Z/2 \times \Z/2.
\end{equation}
We identify the two $\Z/2$-valued global anomalies with the Witten anomalies associated with each $SU(2)$ factor in the Pati-Salam group, a result that follows straightforwardly from Witten's original arguments. We quote the remaining results of our calculations for all bordism groups $\Omega^{\text{Spin}}_{d \leq 5}\left(B(\text{PS})\right)$ in Table~\ref{summary bordism BSM}. 

We note in passing that there are variants on the Pati-Salam gauge group that involve various discrete factors, which complicate the computation of the bordism groups. For example, left-right symmetric models have been proposed in which $G=\text{PS}\rtimes \Z/2$, and there are also models featuring a quotient by a $\Z/2$ subgroup. Unfortunately, neither of the bordism computations for these gauge groups succumb to attack using the simple fibrations considered in this paper.

\subsection{Trinification models}%
\label{sec:Trinification}

In trinification models of grand unification\cite{Glashow:1984gc}, the underlying gauge group is either 
\begin{equation}
\label{eq: trinification gg}
G = SU(3)_C \times SU(3)_L\times SU(3)_R\quad \text{or} 
\quad G = \frac{SU(3)_C\times SU(3)_L\times SU(3)_R}{\Z/3},
\end{equation}
where the $\Z/3$ quotient is the diagonal subgroup of the $(\Z/3)^3$ centre symmetry. In both cases, the SM quarks are packaged into representations $(\bar{\mathbf{3}},\mathbf{1},\mathbf{3})$ and $(\mathbf{3},\bar{\mathbf{3}},\mathbf{1})$, with the leptons transforming in the $(\mathbf{1},\mathbf{3},\bar{\mathbf{3}})$. The model also contains multiple Higgs fields transforming in the $(\mathbf{1},\mathbf{3},\bar{\mathbf{3}})$ representation (each of which contains three SM-like Higgs doublets), needed to break the gauge symmetry down to a SM subgroup;  the first option in (\ref{eq: trinification gg}) is broken down to $G_{\text{SM}}/\Gamma_2$, while the second is broken to $G_{\text{SM}}/\Gamma_6$. Like Pati-Salam models, trinification models are attractive in part because all the gauge, Yukawa, and quartic couplings in the lagrangian can be run to arbitrarily high energies without hitting any Landau poles, thereby exhibiting `total asymptotic freedom'~\cite{Pelaggi:2015kna}.

\subsubsection*{No quotient}
\label{Sec: trin no quotient}

To find out whether there are potential global anomalies when the gauge group is $SU(3)^3$, we compute $\Omega^{\Spin}_{d}\left(BSU(3)^3\right)$.
Since the method is very similar to that used in previous Sections, we will only quote
the results here to avoid repetition. We find
\begin{equation}
\def\arraystretch{1.5}
\arraycolsep=4pt
\begin{array}{c|cccccc}
i & 0 & 1 & 2 & 3 & 4 & 5\\
\hline
\Omega^{\Spin}_i(BSU(3)^3) & \Z & \Z/2 & \Z/2 & 0 & \Z^4 & 0. 
\end{array}
\nonumber
\end{equation}
Since $\Omega^{\Spin}_{5}\left(BSU(3)^3\right)=0$, the trinification models based on this gauge group are free of any global anomalies, regardless of the fermion content.

\subsubsection*{$\Z/3$ quotient}%
\label{Sec: trin Z3 quotient}

\begin{figure}[h]
\includegraphics[width=0.55\textwidth]{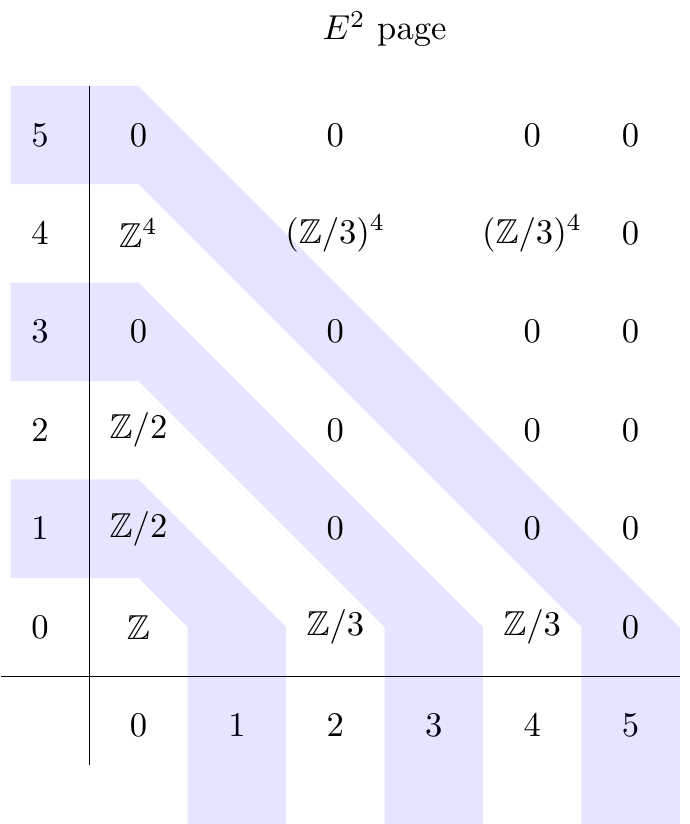}
\centering
\caption{The $E^2$ page  of the Atiyah-Hirzebruch spectral sequence for trinification models featuring a $\Z/3$ quotient of the gauge group.} 
\label{fig: trin E2}
\end{figure}

Now let us consider the option involving a permutation symmetry among the three $SU(3)$ factors, {\em i.e.} where $G =  SU(3)^3/(\Z/3)$. 
We have the fibration $\Z/3\rightarrow SU(3)^3\rightarrow G$, which we can use
the Puppe sequence to turn into the following fibration
\begin{equation}
\label{eq: trinif Puppe}
    BSU(3)^3\longrightarrow BG \longrightarrow B^2(\Z/3) \cong K(\Z/3,2).
\end{equation}
Using this fibration, we can now form the AHSS to find $\Omega_5^{\Spin}(BG)$.
The second page, as we have seen so many times, is given by
\begin{equation}
    E^2_{p,q} = H_p\left(K(\Z/3,2),\Omega_q^{\Spin}(BSU(3)^3)\right)
\nonumber
\end{equation}
which can be constructed using the results for $\Omega_{\text{pt}}^{\Spin}\left(BSU(3)^3)\right)$, which were already calculated in this Subsection. It is displayed in Fig. \ref{fig: trin E2}. One can see immediately that all entries with $p+q = 5$ stabilise already at this page. We can again conclude that
\begin{equation}
    \Omega_5^{\Spin}\left(B\left(\frac{SU(3)_C\times SU(3)_L\times  SU(3)_R}{\Z/3}\right)\right) = 0.
\nonumber
\end{equation}
The other entries with $p+q<5$ also stabilise on this page because all relevant homomorphisms are trivial. The spin bordism groups of lower degrees can be  calculated uniquely apart from  $\Omega^{\Spin}_4$ which involves non-splitting group extensions. It is given by
\begin{equation}
  \Omega^{\Spin}_4\left(B\left(\frac{SU(3)_C\times SU(3)_L\times  SU(3)_R}{\Z/3}\right)\right) = e(\Z/3,\Z^4).
\end{equation}
The full results are given in Table \ref{summary bordism BSM}.

\section{(B)SM theories with  spin$_c$ structures} \label{sec:spin}

Part of the motivation for the bordism-based criterion for anomaly cancellation that we have used in this paper is the desire to define the SM (or our favourite BSM extension) on arbitrary four-manifolds, or at least within some suitable class of four-manifolds. Such a requirement can be motivated by locality, and is certainly a requirement in a quantum theory of gravity in which the geometry (and thus topology) of spacetime cannot be held fixed.

In order to define fermions, one needs to equip spacetime with a spin structure, or a variant thereof with which to stitch together locally-valued spinor fields into globally-defined ones. It is well known that not all orientable four-manifolds admit a spin structure (with $\mathbb{C}P^2$ being a well-known example of an orientable four-manifold that is not spin). The obstruction to being spin is measured by the second Stiefel-Whitney class which takes values in $H^2(\Sigma,\Z/2)$. While $H^2(\Sigma,\Z/2)=0$ for all orientable manifolds in dimension three or fewer, it does not vanish for all four manifolds. One might therefore ask whether the SM and related theories we have explored in this paper can be defined on {\em all} orientable four-manifolds, by not assuming the presence of a spin structure. We invite the reader to consult Appendix \ref{app:spin}, in which we provide more details regarding the definitions of spin structures and the like.

As we noted in \S \ref{sec:bordism}, in the presence of a $U(1)$ gauge symmetry it becomes possible to define spinors using only a spin$_c$ structure on spacetime. The transition functions on a spin$_c$ bundle over an oriented four-manifold $\Sigma$ are valued in the group Spin$_c(4)$, which can be defined by the short exact sequence
\begin{equation} \label{eq:spin c}
0\rightarrow U(1)_A \rightarrow  \text{Spin}_c(4) \rightarrow SO(4)  \rightarrow 0,
\end{equation}
where $U(1)_A$ denotes a gauged symmetry. 
Since all orientable four-manifolds admit a spin$_c$ structure (the obstruction here being in the third Stiefel-Whitney class), one can in principal try to define a four-dimensional gauge theory on all orientable four manifolds by using a spin$_c$ structure. These observations were first made back in 1977~\cite{Hawking:1977ab}, motivated by the authors' desire to define a theory of quantum gravity on all orientable spacetimes.

In order to define all fermions using a spin$_c$ structure, for a particular non-abelian gauge theory (such as one of the SMs), requires there exists a $U(1)$ subgroup of the gauge symmetry, here denoted by $U(1)_A$, such that all fermions in the theory transform in {\em bona fide} representations of the group (\ref{eq:spin c}). Using similar arguments to those given in \S \ref{sec:hypercharge}, this results in constraints on the allowed $U(1)_A$ charges of fermions, which here depend on their {\em spin}. We begin our discussion by recapping what these `spin-charge relations' are, which was recently discussed (in the context of defining similar theories on spin$_c$ manifolds) in Ref.~\cite{Seiberg:2016rsg}.

\subsection{Spin-charge relations}

To derive the spin-charge relations, we require that the SM fermions transform in {\em bona fide} representations of both $\text{Spin}_c(4)$ and $G$, where $G$ is one of the four SM gauge groups listed in Eq. (\ref{SM gauge group}). It is here helpful to write
\begin{equation} \label{spin c 4}
\text{Spin}_c(4) \cong \frac{\text{Spin}(4) \times U(1)_A}{\Z/2} \cong \frac{SU(2)_L \times SU(2)_R \times U(1)_A}{\Z/2},
\end{equation}
A Weyl fermion transforms in the $(\frac{1}{2},0)$ or $(0,\frac{1}{2})$ representation of the $SU(2)_L\times SU(2)_R$ factor. So, when considering Weyl fermions we may restrict our attention to a subgroup of $\text{Spin}_c(4)$ isomorphic to 
\begin{equation} \label{spin c 4 subgroup}
\frac{SU(2)\times U(1)_A}{\Z/2} \cong U(2).
\end{equation}
Thus, by the same argument we used in \S \ref{sec:hypercharge}, one deduces that there exists a normalisation of charges such that all Weyl fermion have {\em odd} charges under $U(1)_A$, in order to define the theory using this spin$_c$ structure.

The question then is, is there any $U(1)_A$ subgroup of $G$ in which all the SM fermions have odd charges? It turns out the answer is no. 
To see why, consider $U(1)_A$ to be generated by
\begin{equation} \label{general u1}
    X = a Y + b \Tilde{T}_3 + c T_3 + d T_8,
\end{equation}
where $Y$ is the generator of hypercharge, 
$$\Tilde{T}_3=
\left( \begin{array}{cc}
1 & 0  \\
0 & -1 \\
\end{array} \right)
$$ is the Cartan generator of (electroweak) $SU(2)$, %and 
$$T_3=
\left( \begin{array}{ccc}
1 & 0 & 0 \\
0 & -1 & 0 \\
0 & 0 & 0 \\
\end{array} \right)
\quad \text{~and~} \quad
T_8=
\left( \begin{array}{ccc}
1 & 0 & 0 \\
0 & 1 & 0 \\
0 & 0 & -2 \\
\end{array} \right)
$$ are the Cartan generators of $SU(3)$ (in a non-standard normalisation which is convenient for our purposes). Eq. \eqref{general u1} defines a general $U(1)_A$ subgroup of $G$. \footnote{Different inclusions of $U(1)$ in the non-abelian factors are related to our choice simply by a change of basis.} 

We then need to decompose all the SM fermion fields into eigenstates of (\ref{general u1}). To wit, consider the left-handed doublet of quarks, $Q$. This needs both an $SU(2)$ index (which we  denote by an upper Greek index $\alpha\in\{1,2\}$) and an $SU(3)$ index (which we  denote by a lower Latin index $i\in\{1,2,3\}$). In this notation, $Q_{\alpha i}$ denotes $2\times 3=6$ Weyl fermions. We thus denote the SM fermion content by the fields $\{ Q^\alpha_i, \; L^\alpha, \; U_i, \; D_i, \; E\}$, which number fifteen in total.

The charges of all the SM fields under the generator \eqref{general u1} are then
\begin{equation} \label{charges}
\begin{array}{c|c}
\textrm{Field} & \textrm{Charge} \\
\hline
Q^1_1 & a+b+c+d \\
Q^2_1 & a-b+c+d \\
Q^1_2 & a+b-c+d \\
Q^2_2 & a-b-c+d \\
Q^1_3 & a+b-2d \\
Q^2_3 & a-b-2d \\
\hline
L^1 & -3a+b \\
L^2 & -3a-b \\
\hline
U_1 & 4a +c+d \\
U_2 & 4a -c+d \\
U_3 & 4a -2d \\
\hline
D_1 & -2a +c+d \\
D_2 & -2a -c+d \\
D_3 & -2a -2d \\
\hline
E & -6a
\end{array}
\\[15pt]
\end{equation}
There are no rational values for $a$, $b$, $c$, and $d$ such that all the charges in this table are odd numbers. To see why, note firstly that the oddness of the charge of $e$ requires that $a=(2n+1)/2$. But then there is no value of $d$ such that both $d_3$ and $u_3$ have odd charge.

We hereby see the restrictiveness of the spin-charge relations: there is in fact no $U(1)$ gauge symmetry in the SM which one can use to define the theory using a spin$_c$ structure. This fact was pointed out in Ref.~\cite{Garcia-Etxebarria:2018ajm}. Hence, given only the gauge symmetries and the fermion content of the SM, one cannot define it on all four-manifolds using a spin$_c$ structure.\footnote{Note that it may still be possible to define the SM consistently on all four-manifolds, but using an even weaker structure than spin$_c$. For example, one may use a spin$-SU(2)$ structure, or a spin$-H$ structure in general where $H$ is any subgroup of $G$. We do not consider such possibilities here.  }

\subsection{Gauging ${B-L}$}

One can instead define a theory on all orientable four-manifolds in which the SM gauge group is extended by an additional $U(1)$ gauge symmetry for which the spin-charge relations are satisfied, such as gauging $B-L$,\footnote{We note that, in this setup, the vector field $A_\mu$ that couples to $B-L$ is not technically an abelian gauge field, since its field strength will not satisfy the Dirac quantisation condition (the corresponding Chern class is only half-integral). Thus, it is not technically correct to describe such a theory as a theory with gauge symmetry $G_{\text{SM}}\times U(1)$. Rather, the vector field $A_\mu$ defines a spin$_c$ connection on $\Sigma$.
} where $B$ is baryon number and $L$ is lepton number. Under $U(1)_{B-L}$ all the SM fermions have odd charges (either $-1$ or $3$), and so this gauge symmetry can be used to define a spin$_c$ structure~\cite{Garcia-Etxebarria:2018ajm}.

Of course, $B-L$ is free of local ABJ-type anomalies. Here we consider global anomalies in SM$\times U(1)$ theories defined on all spin$_c$ manifolds, such as gauged  $B-L$, by computing the bordism groups $\Omega_5^{\text{Spin}_c}(BG)$, for the SM gauge groups listed in Eq. (\ref{SM gauge group}). These bordism groups can be computed using the AHSS associated to a fibration of the form $F\rightarrow BG \rightarrow B$. For example, given the `trivial' fibration $\text{pt} \rightarrow BG \rightarrow BG$, the second page of the AHSS is now 
\begin{equation} \label{AHSS spin c pg 2}
E^2_{p,q}=H_p(B;\Omega^{\text{Spin}_c}_q(F)),
\end{equation}
where the bordism groups of spin$_c$ $q$-manifolds equipped with maps to a point are~\cite{bahri1987eta}
\begin{equation}
\label{eq:bordism-table-spin c}
\def\arraystretch{1.5}
\arraycolsep=4pt
\begin{array}{c|ccccccccccc}
q & 0 & 1 & 2 & 3 & 4 & 5 & 6 & 7 & 8 & 9 & 10 \\
\hline
\Omega^{\text{Spin}_c}_q(\text{pt}) & \Z & 0 & \Z & 0 & \Z^2 & 0 & \Z^2 & 0 & \Z^4 & 0 & \Z^4.
\end{array}
\end{equation}
Interestingly, these groups do not feature any torsion, and moreover they vanish in all odd degrees, at least up to $\Omega^{\text{Spin}_c}_{9}(\text{pt})$.  It then follows immediately that
\begin{equation}
\Omega^{\text{Spin}_c}_{d}(BG_{\text{SM}}) =\Omega^{\text{Spin}_c}_{d}(BG_{\text{SM}}/\Gamma_2)=\Omega^{\text{Spin}_c}_{d}(BG_{\text{SM}}/\Gamma_3)=0 \qquad \text{for all odd~} d \leq 9,
\end{equation}
because non-zero entries in $E^2_{p,q}$ can only appear when $p+q$ is even (since $H_p (BG, \Z)$ also vanishes in all odd degrees for these gauge groups). In particular, these groups vanish in degree $d=5$, so there are no possibilities of global anomalies in any of these theories.

The case where $G=G_{\text{SM}}/\Gamma_6$ is only slightly less straightforward. We may as before proceed via the Puppe sequence to deduce the fibration
\begin{equation}
B(U(2)\times SU(3))\rightarrow B(G_{\text{SM}}/\Gamma_6) \rightarrow K(\Z/3,2),
\end{equation}
and write down the corresponding AHSS, from which one immediately sees that
\begin{equation}
\Omega^{\text{Spin}_c}_{5}(BG_{\text{SM}}/\Gamma_6)=0,
\end{equation}
so again such a theory is automatically free of global anomalies. These conclusions hold when the SM fermion content is extended arbitrarily.

\subsection*{Acknowledgments}
We thank Philip Boyle-Smith for bringing our attention to this problem, as well as Alex Abbott and Oscar Randal-Williams for discussions. We thank Pietro Benetti Genolini, David Tong, Juven Wang, and Edward Witten for helpful comments on the manuscript. BG is partially supported by the STFC consolidated grant ST/P000681/1 and King's College, Cambridge. JD is supported by The Cambridge Trust, and by the STFC consolidated grant ST/P000681/1. NL is supported by the Internal Graduate Scholarship from Trinity College, Cambridge.

\appendix

\section{Spin structures and the like } \label{app:spin}

In this Appendix, we consider fermions defined on a $p$-dimensional smooth spacetime manifold $\Sigma^p$. Fermions are usually defined to be  spinors on $\Sigma^p$. Defining spinors requires a spin structure on spacetime. To explain what a spin structure is, we first assume that $\Sigma^p$ is orientable. A spinor is then a section of a so-called  spinor bundle over $\Sigma^p$, whose structure group is the group $\text{Spin}(p)$, the double cover of $SO(p)$ (which is the structure group of the tangent bundle). What this means is that two locally-valid descriptions of a spinor field, $\Psi_\alpha$ (defined on an open set $U_\alpha$ of $\Sigma^p$) and $\Psi_\beta$ (defined on $U_\beta$), are related by $\Psi_\alpha = T_{\alpha\beta} \Psi_\beta$, for some matrix $T_{\alpha\beta}\in \text{Spin}(p)$ defined on the double-overlap $U_\alpha \cup U_\beta \equiv U_{\alpha\beta}$.\footnote{The spin-valued matrices $T_{\alpha\beta}$ are moreover obtained by lifting the transition functions from the tangent bundle, which are valued in the (orientation-preserving) structure group $SO(p)$. } In order to be able to define spinors globally, we must be able to piece together locally-valid descriptions on open sets $\{U_\alpha\}$ consistently. This requires a set of $\text{Spin}(p)$-valued transition functions defined on every double overlap $U_{\alpha\beta}$, which moreover satisfy a consistency condition on triple overlaps, {\em viz.} $T_{\alpha\beta}\cdot T_{\beta\gamma}\cdot T_{\gamma\alpha}={\bf 1}$ on $U_{\alpha\beta\gamma}$. A consistent set of $\{T_{\alpha\beta}\}$ is called a spin structure on $\Sigma^p$. 

Not every Riemannian manifold admits such a collection of $\text{Spin}(p)$-valued transition functions that satisfy the consistency condition. An orientable manifold admits a spin structure, which can be used to define spinors, if and only if both the first and second Stiefel-Whitney classes  (which take values in $H^1(\Sigma^p,\Z/2)$ and $H^2(\Sigma^p,\Z/2)$ respectively) vanish. If this is the case, $\Sigma^p$ is called a  spin manifold. For example, all orientable manifolds in dimension $p \leq 3$ are spin; whereas four-manifolds are not, necessarily. The $\text{Spin}(p)$-valued $T_{\alpha\beta}$ then define transition functions on a vector bundle $S\rightarrow \Sigma^p$, called a  spinor bundle, of which a fermion field is a section.

This is not the only way to define a geometric object which behaves as a fermion. If spacetime is  non-orientable, alternative structures (called  pin structures) may still be used to define an analogue of the spinor,\footnote{In the unorientable case, the fermion might better be called a `pinor'.} and hence to define fermions. The idea here is very similar to defining spinors in the case that $\Sigma^p$ was orientable, except that now the transition functions of the tangent bundle are valued in $O(p)$, rather than $SO(p)$, because they need not preserve orientation. Consequently, the structure group of the `pinor' bundle is a double cover of $O(p)$, which is called a $\text{Pin}(p)$ group. But now there is not just one such double cover of $O(p)$, but two possible choices called $\text{Pin}^+$ and $\text{Pin}^-$, as follows. One may choose a spatial reflection ${\bf R}$ to satisfy ${\bf R}^2=1$ when acting on spinors, which defines the double cover $\text{Pin}^+$, or choose ${\bf R}^2=-1$, which defines the double cover $\text{Pin}^-$. A pin structure is then defined in a similar way to a spin structure; the $O(p)$-valued transition functions of the tangent bundle are lifted to (say) $\text{Pin}^+$-valued functions, which must satisfy a consistency relation on triple overlaps. A non-orientable manifold that admits a (say) pin$^+$ structure is, not surprisingly, called a pin$^+$ manifold. Again, there are topological obstructions (involving Stiefel-Whitney classes) to defining such pin structures, which are different for pin$^+$ and pin$^-$ structures. Notably, every non-orientable 2-manifold and 3-manifold admits a pin$^-$ structure, but not necessarily a pin$^+$ structure.\footnote{For example, the manifold $\mathbb{R}P^2$ admits only pin$^-$ structures.}

In both the orientable and non-orientable cases, one may in fact still define fermions using weaker structures on $\Sigma^p$, provided there are additional gauge symmetries acting on the fermions. For example, a manifold that is not spin may nonetheless admit a spin$_c$ structure, which is defined analogously to a spin structure, but where the transition functions can be valued in the $\text{Spin}_c(p)$ group rather than $\text{Spin}(p)$. The group  $\text{Spin}_c(p)$ can be defined by the short exact sequence $0\rightarrow U(1) \rightarrow  \text{Spin}_c(p) \rightarrow SO(p) \rightarrow 0$; in an intuitive sense, this ``allows'' the transition functions to vary by a (local) $U(1)$-valued phase, which can be used to ``stitch together'' transition functions where a spin structure might not be possible. If a fermion is acted upon by a $U(1)$ gauge symmetry, then it is invariant under such local $U(1)$ rephasings, and so will be well-defined using only the spin$_c$ structure.
The obstruction to a manifold admitting a spin$_c$ structure now lies in its {\em third} Stiefel-Whitney class valued in $\Z$ (rather than $\Z/2$).  Importantly, all orientable manifolds in dimension $p \leq 4$ are spin$_c$.\footnote{Even `weaker' structures have been used to define fermions on general spacetimes in the quantum gravity literature, using the idea of spin-$G$ structures for various Lie groups $G$~\cite{Back:1978zf,Avis:1979de}. The use of spin-$SU(2)$ structures, for an $SU(2)$ gauge theory, has recently been used to derive a new kind of global anomaly~\cite{Wang:2018qoy}.} Analogously defined pin$_c$ structures may be used to define fermions on non-orientable spacetimes with a $U(1)$ gauge symmetry.

\section{Computation of $H_6(K(\Z/3,2),\Z)$ } \label{App:Serre}

In Ref.~\cite{Garcia-Etxebarria:2018ajm}, a theorem from Ref.~\cite{Tischler1} was used to show that the homology groups $H_i(K(\Z/3,2);\Z)$ are given by
\FloatBarrier
\begin{table}[h!]
  \begin{center}
    \begin{tabular}{C|CCCCCCC}     
      i & 0 & 1 & 2 & 3 & 4 & 5 & 6\\
      \hline
      H^i\left(K(\Z/3,2);\Z\right) & \Z & 0 & \Z/3 & 0 & \Z/3 & 0 & C\times \Z/9 
    \end{tabular}
  \end{center}
\end{table}
\FloatBarrier
\noindent where $C$ is an abelian group of exponent less than or equal to $6$, i.e., the degree of any element in $C$ does not exceed $6$. This means that, {\it a priori}, it has the form
\begin{equation}
  C \cong (\Z/2)^{h_2}\times (\Z/3)^{h_3}\times (\Z/4)^{h_4}\times (\Z/5)^{h_5}\times (\Z/6)^{h_6}
\end{equation}
with $h_i\geq 0$. We will use the Serre spectral sequence to show that $C$ must be of the form
\begin{equation}
  C \cong (\Z/3)^n, \qquad n\geq 0.
\end{equation}

\begin{figure}[h!]
\includegraphics[width=0.6\textwidth]{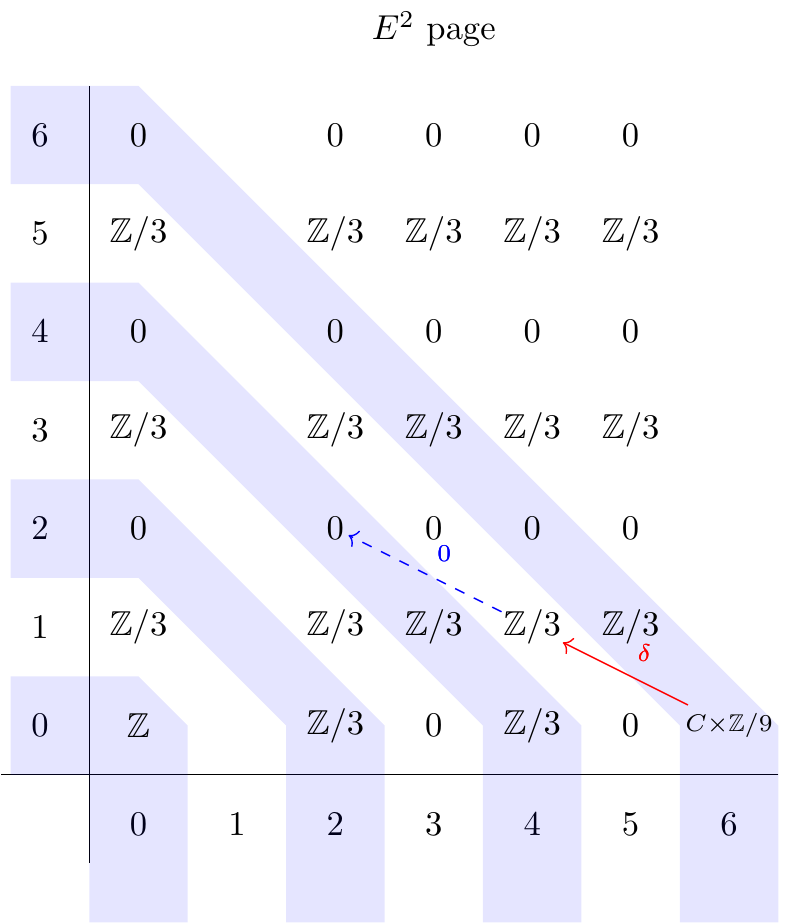}
\centering
  \caption{The $E_2$ page  of the Serre spectral sequence for the fibration (\ref{eq:Serre ss fibration})}
  \label{fig: Serre ss}
\end{figure}

Recall that for a fibration $F\rightarrow X\rightarrow B$, the $(p,q)$ entry on the second page of the Serre spectral sequence is given by \cite{Hatcher:478079}
\begin{equation}
  \label{eq:Serre ss defn}
  E^2_{p,q} = H_p\left(B;H_q\left(F\right)\right)
\end{equation}
The spectral sequence converges to $H_\bullet(X)$, that is, the homology groups of $X$ is determined from the last page of the spectral sequence by\footnote{To be precise, we need to phrase this in terms of filtrations, and solve extension problems to determine the homology groups. However, since the spectral sequence we are interested in converges to $0$ for $p+q>0$, as we shall see momentarily, it follows that all extensions split.}
\begin{equation}
  \label{eq:Serre ss last page}
  H_n\left(X\right) = \bigoplus_{p=0}^nE^\infty_{p,n-p}
\end{equation}
Just like in \cite{Garcia-Etxebarria:2018ajm}, we consider the fibration
\begin{equation}
  \label{eq:Serre ss fibration}
  K\left(\Z/3,1\right)\longrightarrow \star \longrightarrow K\left(\Z/3,2\right)
\end{equation}
where $\star$ is a contractible space. The second page of the Serre spectral sequence is given in figure \ref{fig: Serre ss}.

 Since $H_i(\star) = 0$ for $i>0$, any entry in the Serre spectral sequence apart from $E_{0,0}$ must stabilise to $0$. In particular, the entry $E_{6,0}$ must stabilise to $0$. Since the differential $\delta$ acts trivially on $\Z/2, \Z/4$, and $\Z/5$, these factors would be present in $E^{\infty}_{6,0}$ unless $h_2=h_4=h_5 =0$. 

We can also see that $h_6 = 0$ by a similar argument. Suppose that $h_6\neq 0$. Let $\delta_6$ be a homomorphism from $\Z/6$ to $\Z/3$. There are three choices depending on where it sends the element $1$. The first choice is $\delta_6(1) = 0$, which is the trivial homomorphism, in which case the kernel is $\Z/6$. The second choice and third choice are sending $1$ to $1$ or $2$, both of which result in the same kernel: $\ker \delta_6 \cong \Z/2$. In subsequent pages, the homomorphisms from the $(6,0)$ entry go into either $0$ or $\Z/3$, and can never result in a trivial kernel. Therefore, $E^{\infty}_{6,0} \neq 0$, which is a contradiction. Hence $h_6 = 0$.  This is enough for our purpose: we have determined that 
\begin{equation}
  \label{eq:final form C}
  H_6(K(\Z/3,2),\Z) \cong (\Z/3)^n \times \Z/9, \qquad n\geq 0.
\end{equation}

\section{Two derivations of $\Omega^{\Spin}_5(B(G_{\text{SM}}/\Gamma_3))$}
\label{App: alt Z3}

In this Appendix we give the details of the computation of the spin bordism groups of the SM quotient by $\Z/3$. We present two methods, associated with two different fibrations.

\subsection*{Method 1}

Firstly, we use the AHSS associated to the fibration
\begin{equation}
\text{pt}\rightarrow U(3)\times SU(2) \rightarrow U(3) \times SU(2),
\end{equation}
for which the second page of the AHSS is given by $E^2_{p,q}=H_p (B(U(3)\times SU(2)); \Omega_q^{\text{Spin}}(\text{pt}))$.
The relevant cohomology rings are
\begin{nalign}
  H^\bullet\left(BU(3)\right) =& \Z[c_1,c_2,c_3]\\
  H^\bullet\left(BSU(2)\right) =& \Z[c^\prime_2]
\end{nalign}
where $c_i, c^\prime_i$ are the $i$th Chern classes for $BU(3)$ and $BSU(2)$, respectively. From this, together with the K\"unneth formula in cohomology, we find that $H^2(B(G_{\text{SM}}/\Gamma_3))$ is generated by $c_1$, $H^4(B(G_{\text{SM}}/\Gamma_3))$ by $c^2_1, c_2, c^\prime_2$, and $H^6(B(G_{\text{SM}}/\Gamma_3))$ by $c^3_1,c_1 c^\prime_2, c_1 c_2, c_3$, and again the absence of torsion means these cohomology groups are isomorphic to the corresponding groups in homology.

We again form the AHSS associated to the trivial fibration over a point. The entries on the second page of the AHSS are identical to those of the previous two cases, albeit with different action of the differentials, so we choose not to reproduce the diagram for a third time. Again, the difference to the previous cases shall enter in the action of the differential labelled $\gamma$.

The differentials relevant to the calculation of $\Omega^{\Spin}_4\left(B(G_{\text{SM}}/\Gamma_3)\right)$ and $\Omega^{\Spin}_5\left(B(G_{\text{SM}}/\Gamma_3)\right)$ may be labelled precisely as in Eq. (\ref{differentials SM quotients}) above.
Since $\text{Sq}^2:H^2\rightarrow H^4$ maps $c_1\mapsto c^{2}_1$, we see that both $\alpha, \beta$ maps $\widetilde{c^{2}_1}\mapsto \widetilde{c_1}$ and others to zero, and moreover $\alpha$ maps $2\widetilde{c^{2}_1}$ to zero as before. So we again have $\ker \alpha \cong \Z^3$, $\Ima \alpha \cong \Z/2$, $\ker \beta \cong (\Z/2)^2$, and $\Ima \beta \cong \Z/2$.

We turn to the action of $\gamma$. The relevant Steenrod square is here
\begin{nalign}
\text{Sq}^2:H^4\left(B(G_{\text{SM}}/\Gamma_3);\Z/2\right)&\longrightarrow H^6\left(B(G_{\text{SM}}/\Gamma_3);\Z/2\right)\\
  c^{2}_1 &\mapsto 2c^{3}_1 \equiv 0 \mod\, 2,\\
  c^\prime_2 &\mapsto 0, \\
  c_2 &\mapsto c_1c_2+c_3.
\end{nalign}
So $\gamma$ maps $\widetilde{c_1 c_2}\mapsto \widetilde{c_2}$ and $\widetilde{c_3}\mapsto \widetilde{c_2}$, while mapping other generators to zero. This gives $\Ima \gamma \cong \Z/2$, and hence
\begin{equation}
E^3_{4,1} = \frac{\ker \beta}{\Ima \gamma} \cong \Z/2,
\end{equation}
and this entry stabilises. This is the only non-vanishing entry on the $p+q=5$ diagonal, and so we find
\begin{equation}
\Omega^{\text{Spin}}_5(B(G_{\text{SM}}/\Gamma_3)) \cong \Z/2.
\end{equation}
Since the discrete $\Z/3$ quotient is here embedded `orthogonally' to the $SU(2)$ factor in $G$, we feel safe in suggesting that this $\Z/2$ captures the Witten anomaly coming from the $SU(2)$ factor. As for the previous example, the lower-degree bordism groups are unchanged (see Table~\ref{summary bordism}).

\subsection*{Method 2}

We provide here an alternative proof that $\Omega^{\Spin}_5(B(G_{\text{SM}}/\Gamma_3)) \equiv \Z/2$ using an alternative fibration,
\begin{equation}
  \Z/3 \longrightarrow G_{\text{SM}} \longrightarrow G_{\text{SM}}/\Gamma_3.
  \label{eq: alt SM Z3 fibration}
\end{equation}
After we apply the Puppe sequence, this fibration turns into
\begin{equation}
  \label{eq: alt AHSS Z3 fibration}
  BG_{\text{SM}} \longrightarrow B(G_{\text{SM}}/\Gamma_3) \longrightarrow K(\Z/3,2)
\end{equation}
Using the results for the homology groups of $K(\Z/3,2)$ up to degree 6 given in Appendix \ref{App:Serre}, we can work out the $E^2$ page of the Atiyah-Hirzebruch spectral sequence, given in Figure \ref{fig: AHSS E2 E6  SM/Z3 5}.\footnote{We denote $\Z/m$ by $\Z_m$ in this particular diagram.}  Moreover, we can deduce that the differential $d$ in the $E^6$ page must be trivial, since it is a homomorphism from a product of $\Z/m$ factors with $m$ odd to $\Z/2$. All the entries $E_{p,q}$ with $p+q=5$ now stabilise, and we can read off the spin bordism group as
\begin{equation}
  \Omega^{\Spin}_5(B(G_{\text{SM}}/\Gamma_3)) \equiv \Z/2,
\end{equation}
as claimed.
  \begin{figure}[h!]
   \includegraphics[width=1.0\textwidth]{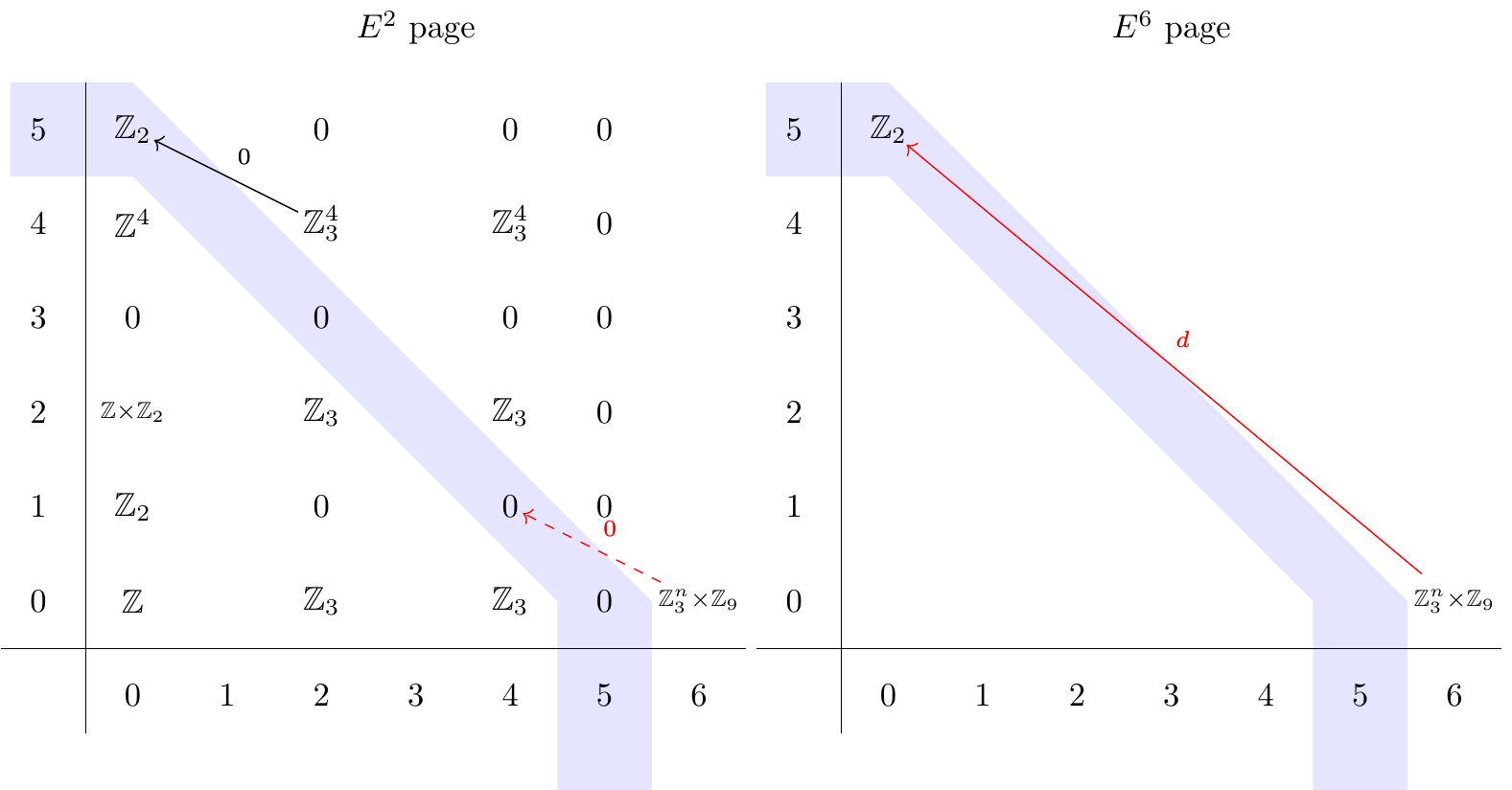} 
\centering
  \caption{The $E^2$ and $E^6$ pages of the Atiyah-Hirzebruch spectral sequence for $G=G_{\text{SM}}/\Gamma_3$ from the fibration (\ref{eq: alt AHSS Z3 fibration}).}
  \label{fig: AHSS E2 E6 SM/Z3 5}
\end{figure}

\section{Decomposing $U(n)$ irreducible representations}
\label{App: U(n) irreps}

The purpose of this Appendix is to decompose an irreducible representation of $U(n) \cong \frac{SU(n)\times U(1)}{\Z/n}$ in terms of the $U(1)$ charge and $ SU(n)$ irreducible representation using character theory, from which we extract the charge constraints presented in \S \ref{sec:hypercharge}.

Let $G$ be a group and $V$ a $d$-dimensional representation of $G$. An element $g\in G$ is represented by a $d\times d$ matrix $R_V(g)$. The character of $g$ in the representation $V$, denoted by $\chi_V(g)$, is defined by
\begin{equation}
  \chi_V(g) = \frac{1}{\text{dim}\,V}\Tr\,R_V(g).
\end{equation}
(We use the normalised character where we have $\chi_V(e) = 1$ for all finite irreducible representation $V$.) From this definition, it is easy to see that the character of $g$ is  a class function, that is, it only depends on the conjugacy class of $g$
\begin{equation}
  \chi_V(g) = \chi_V(hgh^{-1}),\qquad \text{for any}\;h\in G
\end{equation}

We now specialise to the case $G = U(n)$. Since any unitary matrix can be diagonalised by a unitary matrix, any element $g\in U(n)$ is conjugate to a diagonal matrix of the forms
\begin{equation}
  \label{eq: conjugacy class U(n)}
  g \sim \text{~diag~} \left(z_1, z_2, \ldots, z_n\right),\qquad \left|z_i\right| = 1.
\end{equation}
Therefore, a $U(n)$ character can be thought of as a function $\chi^{U(n)}_V: T^n \rightarrow \C$, where $T^n$ is the maximal torus of $U(n)$.

Characters of irreducible representations of $U(n)$ are given by a certain type of symmetric functions called Schur's functions. Let ${\bm \lambda} = \left(\lambda_1, \lambda_2,\ldots, \lambda_n\right)$ be an array of integers satisfying
\begin{equation}
  \lambda_1 \geq \lambda_2 \geq \ldots \geq \lambda_n.
\end{equation}
Note that if $\lambda_n\geq 0$ this is the partition ${\bm \lambda}$ of the non-negative integer $\left|{\bm \lambda}\right| = \lambda_1+\ldots + \lambda_n$. In fact, we can write ${\bm \lambda}$ in terms of an integer $m$ and a {\em bona fide} partition ${\bm \mu} = \left(\mu_1,\ldots, \mu_{n-1}\right)$, with $\mu_i\in \Z$ and 
\begin{equation}
  \mu_1\geq \mu_2 \geq \ldots \geq \mu_{n-1} \geq 0,
\end{equation}
by writing $\lambda_i = m + \mu_i$ for $i=1,\ldots, n-1$ and $\lambda_n = m$. We denote this decomposition by ${\bm \lambda} = (m)^n + {\bm \mu}$. ${\bm \mu}$ can be represented by a Young diagram consisting of $\left|{\bm \mu}\right|$ boxes in total, with $m_i$ boxes in the i\textsuperscript{th} row. We now define Schur's function in $n$ variables ${\bf z} = (z_1,\ldots, z_n)$ by
\begin{equation}
  \label{eq: Schur function}
  s_{{\bm \lambda}}({\bf z}) = \frac{\begin{vmatrix} z_1^{\lambda_1+n-1} & \cdots & z_n^{\lambda_1+n-1}\\
      z_1^{\lambda_2+n-2} & \cdots & z_n^{\lambda_2+n-2}\\
      \vdots & \ddots & \vdots\\
      z_1^{\lambda_n} & \cdots & z_n^{\lambda_n}
    \end{vmatrix}}{\begin{vmatrix} z_1^{n-1} & \cdots & z_n^{n-1}\\
      z_1^{n-2} & \cdots & z_n^{n-2}\\
      \vdots & \ddots & \vdots\\
      z_1^{0} & \cdots & z_n^{0}
    \end{vmatrix}}
\end{equation}
The irreducible characters $\chi^{U(n)}_V({\bf z})$ of $U(n)$ are precisely the Schur functions $s_{{\bm \lambda}}({\bf z})$ \cite{repthy}.

One gets a similar result for the irreducible characters of $\tilde{g}\in SU(n)$. Since $\det \tilde{g} = 1$, it is conjugate to the diagonal matrix of the form
\begin{equation}
  \tilde{g} \sim \text{~diag~}\left(y_1, y_2\,y_1^{-1}, y_3\,y_2^{-1},\ldots, y_{n-1}\,y_{n-2}^{-1}, y_{n-1}^{-1}\right).
\end{equation}
Any irreducible representation of $ SU(n)$ can be labelled by a partition ${\bm \mu}$, and the associated character is given by
\begin{equation}
  \chi^{ SU(n)}_{{\bm \mu}}(y_1,\ldots, y_n) = s_{{\bm \mu}}(y_1, y_2\,y_1^{-1},\ldots, y_{n-1}\,y_{n-2}^{-1}, y_{n-1}^{-1}).
\end{equation}
where $y_i,\; i = 1,\ldots, n-1$ parametrises the maximal torus $\tilde{T}^{n-1}$ of $ SU(n)$.

A $U(n)$ irreducible representation labelled by ${\bm \lambda} = (m)^n + {\bm \mu}$ can be written uniquely in terms of the $ SU(n)$ irreducible representation $V({\bm \lambda})$ and the $U(1)$ charge $q({\bm \lambda})$ as follows.
\begin{equation}
  \label{eq: hpcharge constraint}
  \left( V({\bm \lambda}), q({\bm \lambda})\right) = \left({\bm \mu}, nm + \left|{\bm \mu}\right|\right).
\end{equation}
To see this, we first write $g\in U(n)$ in terms of a $U(1)$ element $e^{i\theta}$ and an element $\tilde{g}\in  SU(n)$ as $g = e^{i\theta} \tilde{g}$. Then the coordinates ${\bf z}$ of $T^n$ is given in terms of $\theta$ and the coordinates ${\bf y}$ of $\tilde{T}^{n-1}$ by
\begin{equation}
  \label{eq: z-y cov}
  z_1 = e^{i\theta}z_1,\quad z_2 = e^{i\theta} y_2\,y_1^{-1},\quad \ldots \quad ,\quad  z_{n-1} = e^{i\theta} y_{n-1}\,y_{n-2}^{-1},\quad z_n = e^{i\theta} y_{n-1}^{-1}.
\end{equation}
In the representation $(q,V)$, $g$ is represented by $e^{iq\theta} R_V(\tilde{g})$. This can be phrased in terms of characters as
\begin{equation}
  \chi^{U(n)}_V\left(z_1,\ldots,z_n\right) = e^{iq\theta} \;\chi^{ SU(n)}_{\tilde{V}}\left(y_1,\ldots, y_{n-1}\right),
\end{equation}
By direct substitution of \eqref{eq: z-y cov} into \eqref{eq: Schur function}, it is easy to show that
\begin{equation}
  \label{eq: U(n) character in U(1)xSU(n) form}
  s_{{\bm \lambda}}({\bf z}) = e^{i\left(nm + \left|{\bm \mu}\right|\right)\theta}\,s_{{\bm \mu}}\left(y_1, y_2\,y_1^{-1},\ldots, y_{n-1}^{-1}\right),
\end{equation}
whence our claim that $(V, q) = ({\bm \mu}, nm+\left|{\bm \mu}\right|)$ follows.

Therefore, for an irreducible representation $({\bm \mu}, q)$ of $SU(n)\times U(1)$ to be a {\em bona fide} irreducible representation of $U(n)$, we need $q$ to be equal to the number of boxes in ${\bm \mu}$ modulo $n$.

This result can be applied to a more complicated scenario. As an example, we consider the group $G = G_{\text{SM}}/\Gamma_6$ which can be realised as  $G = \left(U(3)\times U(2)\right)/U(1)$, where we identify the overall $U(1)$ factor in $U(3)$ with the one in $U(2)$. Our result \eqref{eq: hpcharge constraint} tells us that, for a representation $({\bm \nu}, {\bm \mu}, q)$ of $ SU(3)\times  SU(2)\times U(1)$ to be a {\em bona fide} representation of $G$, we must have
\begin{equation}
  q = \left|{\bm \mu}\right|\;\text{mod}\; 2,\qquad \text{and}\qquad  q = \left|{\bm \nu}\right|\;\text{mod}\; 3.
\end{equation}

\newpage
\bibliography{references}
\bibliographystyle{utphys}

\end{document}